\documentclass[sigconf]{acmart}
\usepackage[utf8]{inputenc}

\usepackage{caption}
\usepackage{graphicx}
\usepackage{subfig}
\usepackage{placeins}
\usepackage{multirow}
\usepackage{booktabs}
\usepackage{tabu}
\usepackage{hyperref}
\usepackage{url}
\usepackage{makecell}
\usepackage{algpseudocode}

\AtBeginDocument{%
  }

\setcopyright{acmcopyright}
\copyrightyear{2018}
\acmYear{2018}
\acmDOI{XXXXXXX.XXXXXXX}
\acmConference[CIKM2022]{}
\acmPrice{15.00}
\acmISBN{978-1-4503-XXXX-X/18/06}

\begin{document}
\title{IRMA: Iterative Repair for graph MAtching}

\author{Barak Babayov}
\affiliation{%
  \institution{Bar-Ilan University}
  \city{Ramat-Gan}
  \country{Israel}
  \postcode{43017-6221}
}
\email{babayov6@gmail.com}

\author{Prof. Yoram Louzoun}
\affiliation{%
  \institution{Bar-Ilan University}
  \city{Ramat-Gan}
  \country{Israel}
  \postcode{43017-6221}
}
\email{louzouy@math.biu.ac.il}

\begin{abstract}

The alignment of two similar graphs from different domains is a well-studied problem. In many practical usages, there is no reliable information or labels over the vertices or edges, leaving structural similarity as the only information available to match such a graph. In such cases, one often assumes a small amount of already aligned vertices - called a seed.
Current state-of-the-art scalable seeded alignment algorithms are based on percolation. Namely, aligned vertices are used to align their neighbors and gradually percolate in parallel in both graphs.  However, percolation-based graph alignment algorithms are still limited in scale-free degree distributions.

We here propose `IRMA' - Iterative Repair for graph MAtching to show that the accuracy of percolation-based algorithms can be improved in real-world graphs with a limited additional computational cost, and with lower run time when used in a parallel version. IRMA starts by creating a primary alignment using an existing percolation algorithm, then it iteratively repairs the mistakes in the previous alignment steps. We prove that IRMA improves on single-iteration algorithms. We then numerically show that it is significantly better than all state-of-the-art seeded graph alignment algorithms on the graphs that they tested.

In scale-free networks, many vertices have a very low degree. Such vertices have a high probability of erroneous alignments. We show that combining iterations with high recall but low precision in the alignment leads in the long run to higher recall and precision for the entire alignment. 
\end{abstract}

\maketitle              

\section{Introduction}
\subsection{Graph Matching/alignment}
In the simplest form of graph matching (GM - often denoted also Graph Alignment), one is given two graphs $G_1$ and $G_2$ known to model the same data (i.e., there is an equivalence between the graphs vertices). For example, $G_1$ may be the friendship graph from the Facebook social network and $G_2$ the friendship graph from the Twitter social network for the same people. In both cases, the vertices are users and there is an edge between two vertices if the corresponding users are friends in the relevant social network. An assumption common to all GM algorithms is that a friend in one network has a higher than random probability of being a friend in the second network. 

The goal of GM is to create a bijection $M: V_1 \rightarrow V_2$, such that $M$ maps a vertex in $V_1$ to a vertex in $V_2$ if they represent the same real-world entities. In the example above, $M$ should connect the profiles in Facebook and Twitter of the same person. We note by $R$ the ground-truth, i.e., the set of all vertices pairs that represent the same entity in both graphs. Given a bijection $M$, if $M(v_1)=v_2$, and $M(v_3)=v_4$, and the edge $(v1,v3) \in E_1$, and $(v2,v4) \in E_2$, the common edge will be defined as a shared edge. The quality measure for the quality of $M$ is usually the \textit{number} of shared edges.

Finding a full bijection is not always optimal, since some vertices may be absent from one of the two graphs. We thus look for a partial bijection: $M: \tilde{V_1}  \subset  V_1 \rightarrow \tilde{V_2}   \subset V_2$, such that the \textit{fraction} of shared edges is maximal. A single edge bijection is obviously a simple solution to that. Thus, a trade-off between the number of shared edges and the fraction of mapped vertices is often required. Such a trade-off is obtained by adding constraints on the number of elements in $\tilde{V_1}$.
\subsection{Seeded Graph Matching}
In the presence of limited initial information on the bijection, one can use seeded GM. In seeded GM, the input contains beyond $(G_1,G_2)$ also a small $seed \subset V_1 \times V_2$, which is a group of vertices pairs known to represent the same real-world entities in $G_1$ and in $G_2$.
A similar problem emerges when the vertices have additional information named labels or meta-data \cite{malhotra2012studying(16), nunes2012resolving(20)}.
For example, the users on Facebook and Twitter may have additional attributes, such as age, address, and user names. It is obvious that a user named ``Bob Marley" in Facebook is much more likely to represent the same person as a user with the same name on Twitter than a user named ``Will Smith". Pairs of vertices with unique and  similar attributes can be used as a seed. Even a very small  seed can simplify the matching.
Here, we focus on seeded GM based, with no labels over the vertices. We propose an iterative approach to seeded GM, which improves the accuracy of existing approaches.
Specifically, we improve the ExpandWhenStuck algorithm (further denoted EWS) \cite{EXPAND} using an Iterative Repair for graph MAtching (IRMA) algorithm. We show that the results are not only better than EWS, but also of all current seeded GM solvers.
\section{Problem Statement}
Formally, we assume two graph $G1=(V1,E1)$ and $G2=(E2,V2)$, with no attributes on vertices or edges (in contrast for example with knowledge graphs - where each edge has a relation type, or labeled graphs, where each vertex has a label or attributes). We further assume a subset $S \subset V_1$ (defined for example as all vertices with a degree above some threshold). Given a bijection $M: V_1 \rightarrow V_2$, where  $M(v_1)=v_2$, and $M(v_3)=v_4$. The edge $(v1,v3) \in E_1$, is shared if $(v2,v4) \in E_2$. We are looking for the bijection maximizing the number of shared edges among all edges, s.t. $v1,v3 \in S$. We focus on real world-graphs with scale-free degree distributions. 
\section{Related Work}
GM is used in different  disciplines. In social networking, GM can be used for  the de-anonymization of data sets from different domains \cite{korula2013efficient(14), yartseva2013performance(30)}. GM is  used on proteins from different species in biology to detect functional equivalences \cite{klau2009new(13), singh2008global(25)}, and to discover a resemblance between images in computer vision \cite{torresani2008feature(26), egozi2012probabilistic(7), wiskott1997face(28)}.

The main progress in GM algorithms is based on machine–learning  methods, and can be broadly divided into two main categories: profile-based and network-based methods. Profile-based methods rely on the vertices meta-data (e.g., username \cite{zafarani2013connecting}, spatio-temporal patterns \cite{riederer2016linking}, posts \cite{goga2013exploiting}, or writing style \cite{narayanan2012feasibility}, etc.) to link accounts across different sites. Network-based methods rely on the graph topological structure \cite{koutra2013big, tan2014mapping, zhou2018deeplink}. In general, machine learning based methods are characterized by a high run time, in both training and deployment. Multiple theoretical bounds for a GM solution were proposed \cite{cullina2016improved, shirani2017seeded, cullina2017exact, kazemi2015can}. They make use of several parameters such as seed size, degree distribution, graphs-overlap, etc. Polynomial run-time complexity algorithms were developed that in exchange for scalability may achieve a good alignment only under certain assumptions \cite{fishkind2012seeded, mossel2020seeded, ding2021efficient, fishkind2019seeded}. GM were also extensively discussed in the context of knowledge graphs, for example for cross-lingual entity alignment or joining  entities that represent the object.

An important aspect of GM is scaling, since its practical use is often in large graphs. Current state-of-the-art scalable seeded GM methods are based on gradual percolation, starting from the seed and expanding through common neighbors. This class of algorithms is referred to as Percolation Graph Matching (PGM) methods \cite{korula2013efficient(14),yartseva2013performance(30), henderson2011s(10)}. Despite having in some cases additional information in the form of labels, \cite{henderson2011s(10)} showed the crucial importance of relying on edges during the process of GM. Both \cite{chiasserini2016social} and \cite{EXPAND} present an improvement to \cite{yartseva2013performance(30)} and are currently among the state-of-the-art PGM algorithms.

We follow here the notations of EWS  with minor changes. In the following text, we refer to the input graphs as $G_1=(V_1,E_1)$ and $G_2=(V_2,E_2)$. When we mention a pair $[u,v]$ or $p$, we always refer to a pair of vertices $[u, v]\in V_1\times V_2$. We denote the pairs $[u,v],[u',v']$ as neighboring pairs if the edge $(u,u')\in E_1$ and the edge $(v,v')\in E_2$. Finally, a pair $[u, v]$ conflicts $M$ if it conflicts an existing pair in $[u',v'] \in M$ (i.e.,  $u=u'$ and $v\neq v'$  or vice versa).

In PGM algorithms, one maintains a score for each candidate pair, and uses the score to gradually build the set of the matched pairs - $M$. 
Adding a constant value to the score (also called the mark) of all neighboring pairs of some given pair is called `spreading out marks'. $marks_t(p)$ is defined as the number of marks that $p$ received from other pairs until time $t$, where the unit of time is the insertion of one pair to $M$.
As there is a high chance for two pairs to have the same amount of marks, we define $score_t([u,v])$ to give a priority to pairs with similar degree:
\begin{equation}
    score_t([u,v]) = marks_t([u,v]) - \epsilon \cdot |d_{1,u} - d_{2,v}|,
\label{eq:score1}
\end{equation}
for an infinitesimal $\epsilon > 0$, where $d_{q,v}$ is the degree of a vertex $v$ in graph $q$. 

In a nutshell, EWS (see code in Figure \ref{fig:ExpandWhenStuck}) starts by adding all the seed pairs into $M$, and spreading out marks to all  their neighbors in the appropriate graphs $G1$, and $G2$. Then, at each time step $t$,  EWS chooses a candidate pair $p' = argmax_p(score_t(p))$ that does not conflict $M$, adds it to $M$ and spreads out marks to all  its neighbors. When there are no pairs left with more than one mark (line 15 in Figure \ref{fig:ExpandWhenStuck}), EWS creates an artificial noisy seed ($A$), and uses it to further spread out marks (line 6 in Figure \ref{fig:ExpandWhenStuck}).
$A$ contains all pairs that are: 1) neighbors of matched pairs 2) do not conflict $M$ 3) never have been used to spread out marks.
The novelty of EWS is the generation of an artificial seed whenever there are no more pairs with more than one mark. The artificial seed is mostly wrong. Yet, EWS manages to use it to match new correct pairs and continue the percolation.
At the end of EWS, the set of mapped pairs $M$ is returned along with $MarkedPairs$ that contains counters of marks for all marked pairs. The last is not needed in EWS,  but is used in IRMA. The  main contribution of EWS is a dramatic reduction in the size of the required seed set for random $G(n,p)$ networks (graph with $n$ vertices and a probability of $p$ for each edge). We here show that in scale free degree distribution, seeded GM algorithms still have a low resolution, and propose a novel method to address that.
\begin{figure}[]
\fbox{\parbox{\linewidth}{
    \underline{\textbf{ExpandWhenStuck}}
	\begin{algorithmic}[1]
        \State $A \leftarrow seed$ is the initial set of seed pairs, $M\leftarrow seed$;
        \State $Z \leftarrow \emptyset $ is the set of used pairs
        \State $MarkedPairs \leftarrow \emptyset $ is the set of all marked pairs along with their number of marks
		\While {($|A| > 0$)}
            \For {all pairs $[u,v] \in A$}
				\State Add the pair [u,v] to $Z$ and add one mark to all of its neighbouring pairs;
			\EndFor
            \While {there exists an unmatched pair with at least 2 marks in $MarkedPairs$ that does not conflict $M$ }
            	\State among those pairs  select the one maximizing $score(p)$;
            	\State Add p=[u,v] to the set $M$;
            	\If {[u,v] $\notin Z$}
            		\State Add one mark to all of its neighbouring pairs and add the 									pair [u,v] to $Z$;
            	\EndIf
            \EndWhile
            \State $A \leftarrow $ all neighboring pairs [u,v] of matched pairs M s.t.  [u,v]$\notin Z$, $u \notin V_1(M)$ and $v \notin V_2(M)$;
		\EndWhile\\        
        \Return $M$, $MarkedPairs$;
	\end{algorithmic}
}}
	\caption{ExpandWhenStuck main algorithm}
	\label{fig:ExpandWhenStuck}
\end{figure}
\section{Novelty}
Most seeded GM algorithm relay on mark spreading from known pairs to unknown pairs. Such algorithms rely on two basic assumptions. First, most true pairs have a large enough number of neighbors that can pass them marks. Second, false pairs have a low probability of receiving a large number of marks. Both assumptions fail in scale-free degree distribution networks. Random pairs of very high degree vertices will receive a lot of marks. In contrast,  low degree true pairs will receive practically no marks. IRMA proposes solutions to these two limitations:
\begin{itemize}
    \item IRMA contains an iterative approach to correct for erroneous pairing. We show that even if a pair of vertices received a high number of marks. Over time, it will receive less mark on average than true pairs. Thus, an iterative approach is proposed to correct erroneous matching of vertices.
    \item In order to reach low degree vertices, IRMA has an Expansion/Exploration phase allowing to pair low degree pairs, even with a low accuracy. We show that such a phase increases on the long run both the accuracy and the recall.
\end{itemize}
We show that these two elements drastically improve the performance of seeded GM.
\section{Methods}
\subsection{Evaluation Methods and Stopping Criteria}
\label{sec:Evaluation}
We use precision and recall to evaluate the performance of algorithms: (i) Precision refers to the fraction of true matches in the set of matched vertices (i.e., pairs in $M$ that are in $R$) - $Precision = \dfrac{|M \cap R|}{|M|}$, and (ii) Recall is the size of the intersect of $M$ and $R$ out of the size of $R$  - $Recall = \dfrac{|M \cap R|}{|R|}$, where $R$ is the set of all pairs of vertices that represent the same entity in both of the graphs. To compare the performance of GM algorithms, we also report the  F1–score.
To approximate the quality of a solution during the run time, assuming no known ground truth - namely, given an input to the seeded-GM problem and two possible maps $M,M': V_1 \rightarrow V_2$, we use $weight(M)$ as a score to compare the quality of their mapping - $Weight(M) = |\{(u,v) | (u,v)\in E_1, (M(u),M(v)) \in E_2 \}|.$
\subsection{Data Sets}
\label{sec:Data_Sets}
For fully simulated graphs, we use the above sampling method over  $G(n,p)$ Erdos-Renyi graphs \cite{Erdos-Renyi}, defined as a graph with $n$ vertices, where every edge of the possible $n\choose 2$ exists with a probability of $p$. It is common to mark two graphs created by sampling from an Erdos-Renyi graph as $G_1,G_2 = G(n,p,s)$ \cite{pedarsani2011privacy}. 
To test our algorithm on graphs that better represent real-world data, yet to control their level of similarity, we used sampling over the following real-world graphs (further denoted  as graph 1, graph 2, and so on, according to their order here (Table \ref{table:data_sets})).
\begin{table}[h]
\begin{center}
\begin{tabular}{|c|c|c|c|p{1cm}|} 
\hline
Number & Name & Nodes & Edges & Average degree \\
\hline
1 & Fb-pages-media & 27,900 & 206,000 & 14 \\
\hline
 2 & Soc-brightkite &56,700 & 212,900& 7.8 \\
\hline
 3 & Soc-epinions & 26,600 & 100,100& 7.9 \\
\hline
4 & Soc-gemsec-HU  &47,500 & 222,900 & 9.4 \\
\hline
5 & Soc-sign-Slashdot081106 & 77,300 & 516,600 & 12.1 \\
\hline
6 & Deezer\_europe\_edges & 28,300 & 92,800 & 6.6 \\
\hline
\end{tabular}
\end{center}
\caption{Real-world data set graphs.}
\label{table:data_sets}
\end{table}
\begin{enumerate}
\item Fb-pages-media  - Data collected about Facebook pages (November 2017). 
(http://networkrepository.com/fb-pages-media.php).
\item   Soc-brightkite - Brightkite is a location-based social networking service provider where users shared their locations by checking-in. 
(http://networkrepository.com/soc-brightkite.php).
\item Soc-epinions - Controversial Users Demand Local Trust Metrics: An Experimental Study on epinions.com Community (http://networkrepository.com/soc-epinions.php).
\item Soc-gemsec-HU - The data was collected from the music streaming service Deezer (November 2017). These datasets represent friendship graphs of users from 3 European countries.
(http://networkrepository.com/soc-gemsec-HU.php).
\item Soc-sign-Slashdot081106 - Slashdot Zoo signed social network from November 6 2008. It is noteworthy that this graph was also used in \cite{EXPAND} (http://networkrepository.com/soc-sign-Slashdot081106.php).
\item Deezer\_europe\_edges - A social network of Deezer users which was collected from the public API in March 2020. 
(http://snap.stanford.edu/data/feather-deezer-social.html).
\end{enumerate}
\subsection{IRMA}
\label{sec:Iterative_approach}
The EWS is greedy. At step $t$, it adds to $M$ the pair maximizing $score_t(p)$. If at time $t$ the algorithm chooses some wrong pair $[u',v]$ that obeys $score_t([u',v]) > score_t([u,v])$ for the correct pair $[u,v]$,  $[u,v]$  may receive many marks from neighbors later in the algorithm, such that $score_\infty([u',v]) < score_\infty([u,v])$ . We argue that such a scenario is highly likely, yet, $[u,v]$ will never get into $M$ as it conflicts with another pair already in $M - [u',v]$.

We thus propose to use the  scores at the end of the algorithm  ($score_\infty([u',v]) < score_\infty([u,v])$) to improve the matching. We suggest an iterative improvement of the match, using the final score of the previous iteration. $mark_{i,t}(p)$ is defined as the number of marks gained by pair $p$ during the $i$-th iteration until time $t$; $score_{i,t}(p)$ is defined accordingly as 
\begin{equation}
    score_{i,t}([u,v]) = marks_{i,t}([u,v]) - \epsilon * |d_{1,u} - d_{2,v}|.
\label{eq:score_2}
\end{equation}
Based on these scores, one can establish:
\begin{equation}
    \bar{mar}k_{i,t}(p) = max(mark_{i,t}(p),mark_{i-1,\infty}(p)),
\label{eq:mark_max}
\end{equation}
and,
\begin{equation}
    \bar{score}_t([u,v]) = \bar{mar}k_{i,t}([u,v]) - \epsilon * |d_{1,u} - d_{2,v}|.
\label{eq:score_3}
\end{equation}
IRMA starts with the initialization of $M$ to be the seed set, then it performs a standard EWS iteration (See pseudo code for the Repairing-Iteration and IRMA in Figures \ref{alg:RepairingIteration} and \ref{alg:IRMA}, respectively). In the following iterations, at time $t$, while there is a candidate pair with $\bar{mar}k_t(p)>1$, IRMA adds to $M$ the pair $p$ maximizing $\bar{score}_t(p)$ and does not conflict $M$, and spread marks out of it.

IRMA stops the iterations when the mapping quality stops increasing.  Formally, the $i$-th iteration starts by initializing $M=seed$, then, at time $t$, it adds to $M$ the candidate pair obeying:
$ p=argmax_p\{\bar{score}_{i,t}(p)\}$,
and spread marks out of it - updating $mark_{i,t}$. The iteration ends when no candidate pair $p$, that does not conflicts $M$, satisfies: $\bar{mar}k_{i,t}(p)>1$.

Since in real-life scenario, we do not know the real mapping, we use $weight(M)$ to estimate the quality of the score at the current iteration (Section \ref{sec:Evaluation}). We compute $weight(M_i)$ $\forall i$ where $M_i$ is the matching at the end of the $i$-th iteration. Whenever $weight(M_i) \leq weight(M_{i-1})$, IRMA stops and returns $M_{i-1}$. In practice, IRMA stops when $weight(M) \leq (1+ \delta)*weight(M_{i-1})$, where $\delta$ was empirically set to $0.01$. Note that this does not ensure convergence of the mapping, only of its score.
\begin{figure}[h]
\fbox{\parbox{\linewidth}{
    \underline{\textbf{Repairing Iteration}}
	\begin{algorithmic}[1]
        \State $MarkedPairs_i$ is an input of all marks from the previous iteration
        \State $MarkedPairs_{i+1} \leftarrow \emptyset$, $M\leftarrow A_0$;
        \State Spread out marks from all pairs in $M$
        \While {there exists an unmatched pair with at least 2 marks in $MarkedPairs_i$ or $MarkedPairs_{i+1}$ that does not conflict  $M$ }
        \State Among those pairs select the pair $p = argmax_p\{\bar{score}_t(p)\}$;
        \State Add $p$ to the set $M$;
        \State Add one mark in $MarkedPairs_{i+1}$ to all of its neighboring pairs and add the $p$ to $Z$ 
        \EndWhile\\
        \Return $M, MarkedPairs_{i+1}$
	\end{algorithmic} 
	}}
	\caption{Repairing Iteration receive the marks from the previous iteration along with the seed as input and builds a map taking into consideration the marks gained at the previous and current iteration.} 
	\label{alg:RepairingIteration}
\end{figure}
\begin{figure}[h]
\fbox{\parbox{\linewidth}{
	\underline{\textbf{IRMA}} (Iterative Repair for graph MAtching)
	\begin{algorithmic}[1]
      \State $M, MarkedPairs \leftarrow ExpandWhenStuck(A_0)$
      \State $M_0 = \emptyset$
      \While{ $weight(M) > (1+ \delta)*weight(M_0)$ }
        \State $M_0 = M$
        \State $M, MarkedPairs \leftarrow Repairing Iteration(MarkedPairs)$
      \EndWhile\\
      \Return $M$
	\end{algorithmic} 
	}}
	\caption{IRMA builds a primarily map using EWS and repeatedly improves it by running `Repairing Iteration'. It uses $weight(M)$ as an indication for convergence by the stop condition that appears in line 3.} 
	\label{alg:IRMA}
\end{figure}
\subsection{Expansion/Exploration Boost}
Each IRMA iteration ends, when there are no pairs left with at least $2$ marks, that do not conflict $M$. This threshold is a trade-off between the precision and the recall. If one sets the threshold to a high value, $M$ will only contain pairs with high confidence, yet the percolation will stop early - leading to high precision but low recall. Similarly, setting the threshold to one will increase recall at the expense of precision.

However, in scale free distribution, many vertices have a low degree and mya never be reached. We thus  suggest performing a ``exploration iteration" with a threshold of $1$ mark, once IRMA converged. This leads to a drop in  precision, but an increase in the recall. Then, one can run regular "repairing" iterations again to gradually restore the precision.
The idea is that while wrong pairs could not comply with the next iterations' threshold of $2$ marks, correct pairs might lead to unexplored areas of the graphs. Such pairs gain marks by their newly revealed neighbors, \textit{a-posteriori} justifying their insertion to $M$.
\subsubsection{Break Condition}
\label{sec:Break_Condition}
Since the iterations performed after the `noisy iteration' are meant to filter out pairs with lower certainty, we can no longer expect $|weight(M)|$ to increase between iterations. In practice, the second stage of IRMA (after the expansion boost), restores the precision within a few iterations and rapidly stops improving $M_i$. We thus empirically set the IRMA to always stop after four repairing iterations, after the expansion boost.
\subsection{Parallel Version}
\label{sec:Parallel_Version}
In order to develop a GPU version of IRMA, we propose a parallel version. The bottleneck of EWS is in spreading out marks from $[u,v]$, which costs $deg_1(u)*deg_2(v)$ updates to the queue of marks. Ideally, we would like to perform multiple mark spreading steps in parallel. However, this  impossible since the pair chosen at time $t$ depends on the marks spread out in the previous step. Section 6.3 in \cite{EXPAND}  presents a paralleled version where the main loop has been split into epochs. This version of EWS starts by spreading out marks from the seed, then at each epoch the algorithm greedily takes all possible pairs from the queue one by one - without spreading any mark. When the queue is eventually empty, it simultaneously spread marks from all pairs selected at the current epoch, creating the queue to the next epoch. This method has the advantage of being extremely fast, allowing input graphs with millions of vertices, and has been argued not to fundamentally affect the performance of the algorithm.

We expand this logic to parallelize IRMA.
The $i$-th iteration gets $queue_{i-1}$ as input and starts by greedily adding all possible pairs from the queue into $M_i$ one by one - without spreading any mark. Then, when the queue is  empty, we simultaneously spread out marks from all pairs in $M_i$ creating $queue_i$. Iteration $i$ returns $M_i$ and $queue_i$.
\subsection{Experimental Setup}
For each graph (See Section \ref{sec:Data_Sets}), we use different graphs-overlap ($s$) values in the range of $[0.4-0.8]$, and various combinations of seed size, and $s$ value to test that our results are robust to the graph choice, the value of $s$ and the seed size.

When comparing the performance along with different seed sizes, we start with seed sizes of $25, 200, 50, 150, 25$ and $60$ respectively for graphs $1-6$. The first seed size is then multiplied by $2,4,8$ and $16$ (see figures \ref{fig:graph2_by_seed}, \ref{fig:IRMA}, \ref{fig:parallel_main_plots}). We report for each graph the precision, the recall or the F1 score for each realization. There is no division to training and test, nor parameter hypertuning, since there are no free parameters, and no ground truth used except for the seed in IRMA.

To examine the performance of seeded GM algorithms, one needs a pair of graphs, where at least a part of the vertices correspond to the same entities. Given $G=(V,E)$ and $s$, we generate $G_1=(V_1,E_1)$, and  $G_2=(V_2,E_2)$ by twice randomly and independently removing edges $e \in E$ with a probability of $1-s$. We then remove vertices with no edges in $G_1$ and in $G_2$ separately. The edge overlap between $G_1$ and$G_2$ increases with $s$. We name $s$  the `graphs-overlap'.
\section{Results}
\subsection{EWS on Real-World Graphs}
IRMA can be applied to any percolation algorithm. We here use  EWS \cite{EXPAND} for each iteration, but any other algorithms can be used. We  first tested the accuracy of EWS on a set of real-world graphs (section \ref{sec:Data_Sets}).
We use each of those graphs to create two partially overlapping graphs by twice removing edges from the same graph. The fraction of edges from the original graph ($G$) maintained in each of the partially overlapping graphs $G_1, G_2$ is denoted $s$. We use values of $s$ in the range of $[0.4,0.8]$.

\begin{figure*}[h]
    \centering
     \subfloat a.{%
        \includegraphics[width=0.33\linewidth]{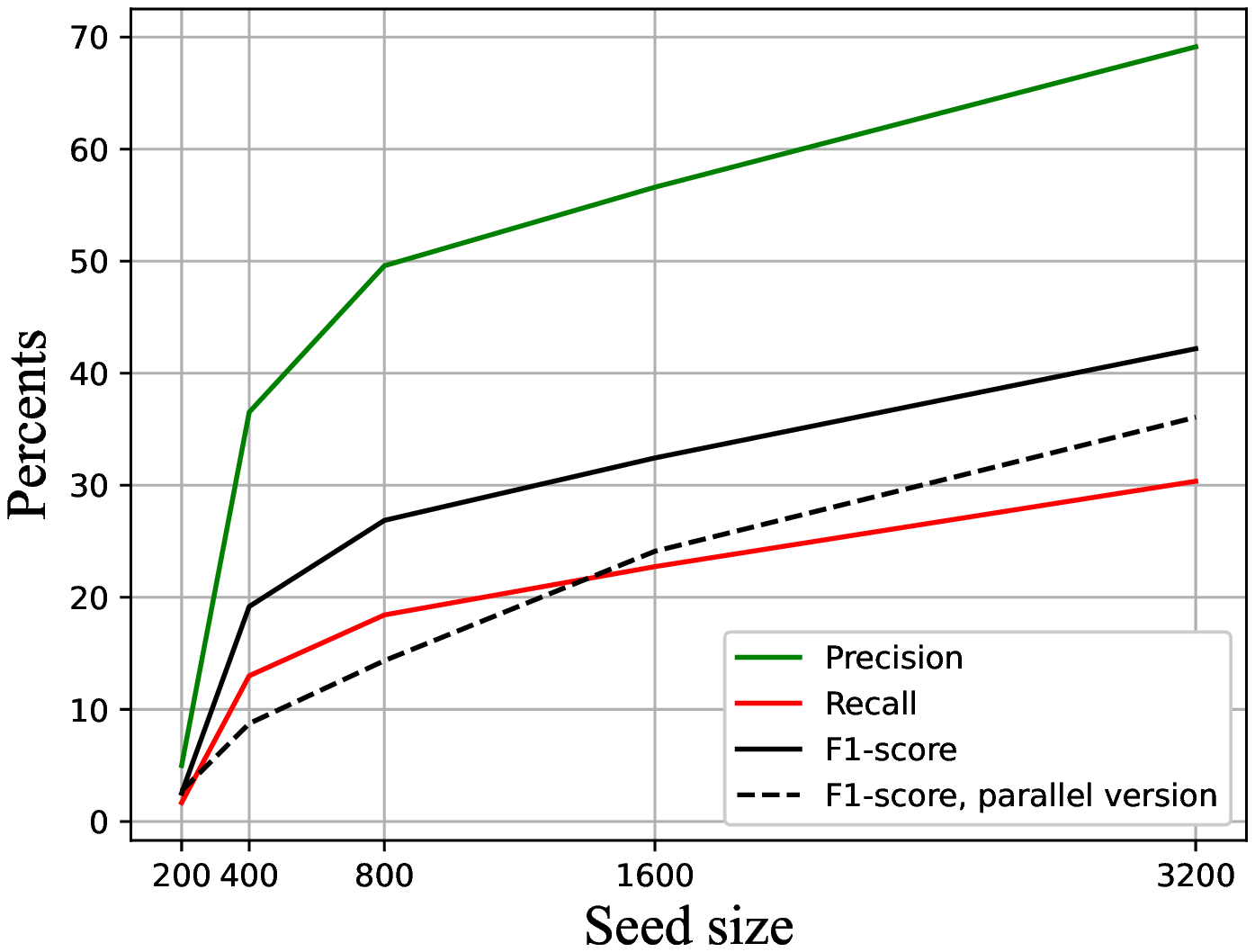}%
        }%
     \subfloat b.{%
        \includegraphics[width=0.33\linewidth]{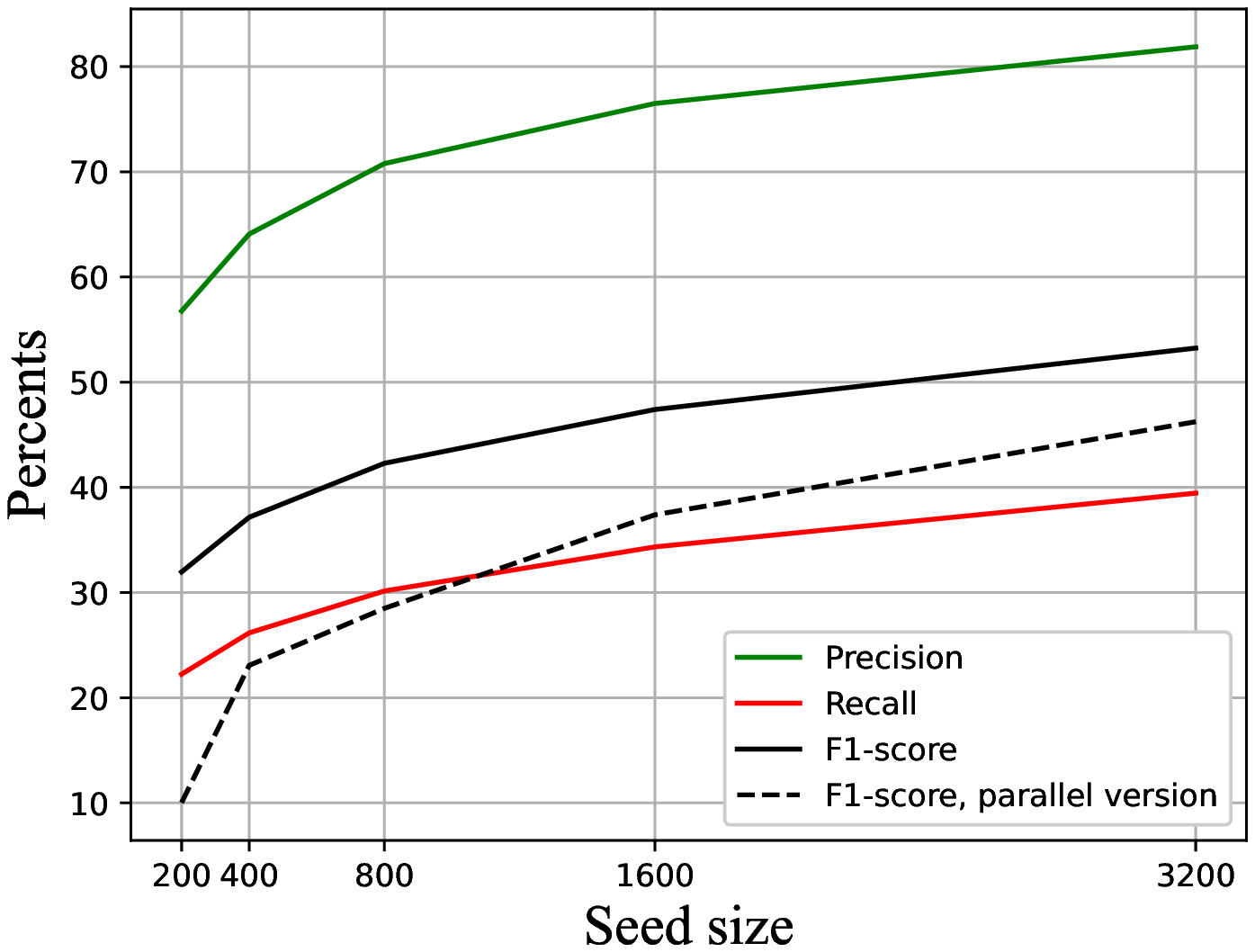}%
        }%
    \subfloat c.{%
        \includegraphics[width=0.33\linewidth]{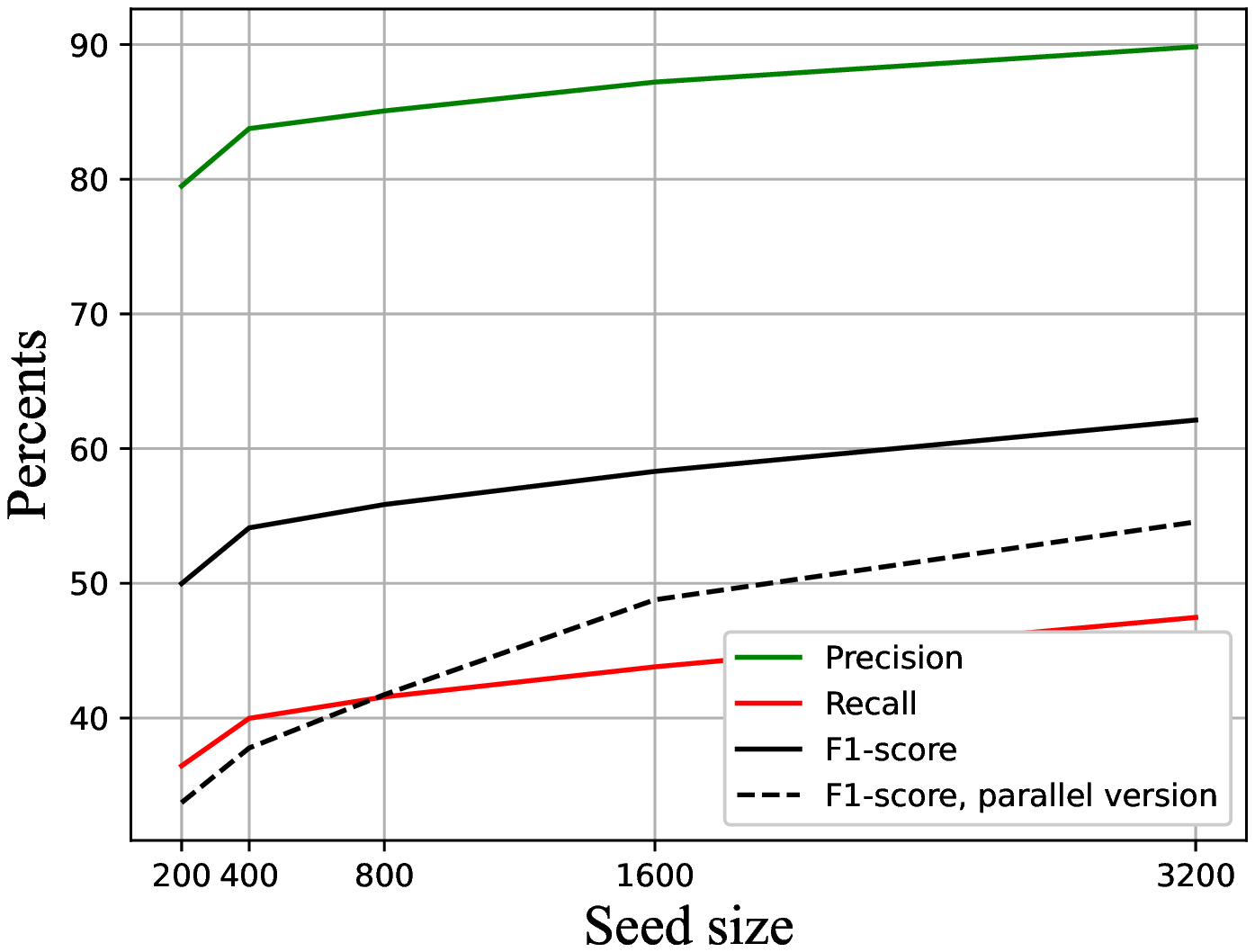}%
        }%
   \caption{ExpandWhenStuck algorithm: Precision, recall, F1-score and F1-score for the parallel version over graph 2 as a function of seed size. Sub-figures a-c use increasing $s$ values of $0.5,0.6,0.7$, respectively. Even with large seed size, performances are significantly worse than the ones originally reported in  $G(n,p)$ Erdos-Renyi graphs. In addition, there is a consistent large gap between recall and precision.}
   \label{fig:graph2_by_seed}
\end{figure*}

We computed the precision, recall and F1-score of EWS over all the  graphs  with graphs overlap ($s$) value of $0.5,0.6,0.7$  as a function of seed size (Figure \ref{fig:graph2_by_seed} a-c, respectively for graph 2, Other graphs are similar). Lower values of $s$, produced recall values of less than 0.3 even for a very large seed.
All accuracy measures are highly sensitive to $s$. The reason is that a correct pair $[u,v]\in G_1 \times G_2$, corresponding to vertex $w \in G$ with degree $d$, has an expected number of $s^2d$ common neighbors, which is approximately the number of marks it will get. In random $G(n,p)$ graphs, most vertices have similar degree (a normal distribution of degrees), and if $s^2*E(d)>2$, EWS typically works. More precisely, \cite{EXPAND} present a simplified version to the EWS and show that the high accuracy in $G(n,p,s)$ requires an average degree of above  $O(\dfrac{log(n)}{s^2})$. 
However, in scale free degree distributions, the vast majority of vertices obey $d<E(d)$. As such, low graphs-overlap values prevent most vertices from gaining enough marks. 

The precision (and also recall and F1) is typically a monotonic non-decreasing function of the seed size, with no clear threshold (Figure. \ref{fig:graph2_by_seed}). The ratio between the precision and the recall as defined here are equal to the ratio between $|M|$ and $G$, i.e., the algorithm fails to percolate through the entire graph.  Again, in scale free  degree distributions, many low-degree vertices will never collect enough marks. As a result, $M$ often contains only part of the possible pairs to match.
\subsection{Parallel Version of EWS}
The parallel version of EWS splits each iteration into epochs, in which it adds to $M$ every possible candidate pair without spreading out marks, and then spreads out marks simultaneously for the next epoch (Figure \ref{fig:precision_and_recall} for some graphs. Other graphs are similar).
The parallel version requires a larger seed in order to achieve  results similar to the original version. Together with the former findings, we can conclude that while  EWS has an excellent accuracy in $G(n,p)$, it has two main limitations: A) Significantly lower performances on real-world graphs than in $G(N,p)$, and difficulty to handle graphs-overlap lower than $0.7$ in real-world networks. B)  A large difference in real-world networks between the parallel and sequential versions. 

We here propose that an iterative version of EWS can reduce these limitations.
    

\subsection{IRMA Algorithm}
IRMA is an iterative Graph Matching solver (section \ref{sec:Iterative_approach}). We first run EWS, and then perform repairing iterations. In each iteration, IRMA uses the marks received at the end of the former iteration to build a new and better map (Figures \ref{alg:RepairingIteration} and \ref{alg:IRMA}). The intuition to keep adding recommendations to mapped pairs and use them on the next iteration is simple. Correct pairs share more common neighbors than the wrong ones in the average case, therefore correct pairs are more likely to gain more recommendations, even if they initially contradict another assignment. 
\subsection{IRMA improves accuracy along iterations}
$score_t(p)$ is defined by $marks_t(p)$ (the number of marks $p$ gained until time $t$) and uses the difference in the vertices degree to break ties (Eq. \ref{eq:score1}). While the often in real time a wrong pairing obtains a higher score than a real pairing, consistently, the score of wrong pairs at infinity (or the end of the analysis) is lower (Data not shown).
\subsubsection{Proof of improvement}
We demonstrate the advantage of IRMA over EWS by showing that performing even one iteration of a `Parallel Repairing Iteration' lowers the probability of mapping wrong pairs when running over $G(n,p,s)$ graphs. 
$Score_{i,t}(p)$ is based on $marks_{i,t}(p)$ (the number of marks $p$ gained during the $i$-th iteration until time $t$) see Eq. \ref{eq:score_2}. Intuitively, if $score_{1,\infty}(p)$ is better than $score_{1,t}(p)$, the marks received during the second iteration should be more accurate than in the first iteration, improving $score_{2,t}(p)$, therefore improving $\bar{score_t}(p) =$ $max(score_{1,\infty}(p),score_{2,t}(p))$. As argued by \cite{EXPAND} when analyzing their algorithm, EWS has a complex property of generating noisy seed whenever it gets stuck, which is hard to analyze. For that reason, we  analyze a simpler version of that algorithm - ExpandOnce where an artificial seed is generated only at the beginning of the algorithm to ensure a smooth percolation. In addition, When a pair crosses the minimum threshold of marks, it immediately is inserted into $M$ (assuming it is not conflicting an existing pair). ExpandOnce keeps percolating by randomly choosing one pair at time that has not been choose before, and uses it to spread out marks. 
\\
\\
\textbf{Theorem 1}: Given $G_1, G_2 \leftarrow G(n,\theta,s)$, let $p' = [u,v']$ be a wrong pair inserted into $M$ at time $t$ and let $p=[u,v]$ be a right pair conflicting $p'$. Assuming at some time $\bar{t} > t$ a correct pair has been used to spread out marks, the following applies:
\begin{multline}
    \mathbb{E}[score_{\infty}(p) - score_{\infty}(p')] > \mathbb{E}[score_{t}(p) - score_{t}(p')]
\label{eq:proof1}
\end{multline}
Whenever we choose a wrong pair $[u,v']$ before the right pair $[u,v]$, we will eventually avoid the latter insertion of $[u,v]$, since it will conflict $M$. Theorem 1 suggests that using $score_\infty$ will reduce the probability of a mistake in the next iteration - eventually reducing the number of wrong pairs in $M$.\\
\textbf{Proof of Theorem 1}:\\
Eq. \ref{eq:proof1} in Theorem 1 is equivalent to:
\begin{multline}
\mathbb{E}[score_{\infty}(p) - score_{t}(p)] > \mathbb{E}[score_{\infty}(p') - score_{t}(p')]
\label{eq:proof2}
\end{multline}
In other words, the expected number of marks that $p$ will get from now on, will be higher than the expected number of marks  $p'$ will receive. Let $p_{t'}=[\alpha,\beta]$ be a pair used to spread out marks at time $t'>t$: 
\begin{itemize}
    \item If $p_{t'}$ is a correct pair, $\alpha$ and $\beta$ represent the same vertex $\gamma \in V$ ($u$ and $v$ represented by the same vertex $w \in V$ as well). $p=[u,v]$ gets one mark if there are edges $(\alpha,u)\in E_1$ and $(\beta,v)\in E_2$. This requires the edge $(\gamma, w)$ to exist in $E$ (which happens with probability $\theta$) and to be sampled in $E_1, E_2$ (happens with probability $s^2$). On the other hand, $p'$ gets a mark if $(\alpha,u)\in E_1 and (\beta,v')\in G_2$. This requires two different edges to exist in $E$ (probability of $\theta^2$) and to be sampled to $E_1$ and $E_2$ accordingly (probability of $s^2$).
    \item If $p_{t'}$ is a wrong pair, $\alpha$ and $\beta$ are represented by the different vertices $\alpha',\beta' \in V$, (note that $p_{t'}$ cannot conflict $p'$). The pair $p$ gets one mark if there are edges $(\alpha,u)\in E_1$ and $(\beta,v)\in E_2$. This happens with probability $\theta^2s^2$. The pair $p'$ gets one mark if there are edges $(\alpha,u)\in E_1$ and $(\beta,v')\in E_2$, which also happens with probability $\theta^2s^2$. In fact, it is possible that one of the pairs $p_{t'}$ will have the form $[k,v]$. In that case, $p'$ gets a mark with probability $\theta^2s^2$ while $p$ gets none, as $[u,v']$ can never be a neighboring pair of $[k,v']$.
\end{itemize}

Let us denote by  $\Lambda_t$ the number of right pairs used to spread out marks until time $t$, and by $\Psi_t$ be the number of wrong pairs used to spread out marks until time $t$.
Let us further denote $\bar{\Lambda}_t = \Lambda_\infty - \Lambda_t$ and similarly $\bar{\Psi}_t = \Psi_\infty - \Psi_t$. Using the analysis above, one can compute:
\begin{multline}
    \mathbb{E}[score_{\infty}(p) - score_{t}(p)] \geq \bar{\Lambda}_t * s^2\theta + (\bar{\Psi}_t -1) * s^2\theta^2
\label{eq:proof3}
\end{multline}
\begin{multline}
    \mathbb{E}[score_{\infty}(p') - score_{t}(p')] = \bar{\Lambda}_t * s^2\theta^2 + \bar{\Psi}_t * s^2\theta^2.
\label{eq:proof4}
\end{multline}
Combining Eq. \ref{eq:proof3} and \ref{eq:proof4}, we obtain:
\begin{multline}
    \mathbb{E}[score_{\infty}(p) - score_{t}(p)] \geq \bar{\Lambda}_t * s^2\theta(1-\theta) -s^2\theta^2 \\ +\mathbb{E}[score_{\infty}(p') - score_{t}(p')]
\label{eq:proof5}
\end{multline}
To prove Eq. \ref{eq:proof2}, it remains to show that $\bar{\Lambda}_t * s^2\theta(1-\theta) -s^2\theta^2 > 0$. Since we assumed that some correct pair has been used to spread out marks at time $\bar{t} >t$, we have $\bar{\Lambda}_t\geq 1$. $\theta$ is the probability of an edge to  exist, and as we only focus on sparse graphs we assume $\theta<<0.5$, leading to:
\begin{multline}
    \bar{\Lambda}_t * s^2\theta(1-\theta) -s^2\theta^2 = (\bar{\Lambda}_t-1) * s^2\theta(1-\theta)+ s^2\theta(1-\theta) -s^2\theta^2 \\
    = (\bar{\Lambda}_t-1) * s^2\theta(1-\theta)+ s^2\theta (1-2\theta) > 0 
\label{eq:proof6}
\end{multline}
\hfill $\square$
\subsection{Experimental validation}
IRMA builds $M_i$ during the $i$-th iteration using $$\bar{score}_{i,t}(p) = max(score_{i,t}(p),score_{i-1,\infty}(p).$$
It then uses  the advantage of each iteration over the previous one. We have shown the expected advantage of the first Repairing-Iteration over EWS, so $score_{2,\infty}(p)$ is more accurate than $score_{1,\infty}(p)$ on average. The proof above did not use any assumption on the way the initial marks were produced. Thus, the same logic should apply to all following  iterations. Thus,  $\bar{score}_{3,t}(p)$ is expected to be more accurate than $\bar{score}_{2,t}(p)$, and the same holds for all following iterations.

\begin{figure*}[h]
    \centering
    \subfloat a.{%
        \includegraphics[width=0.33\linewidth]{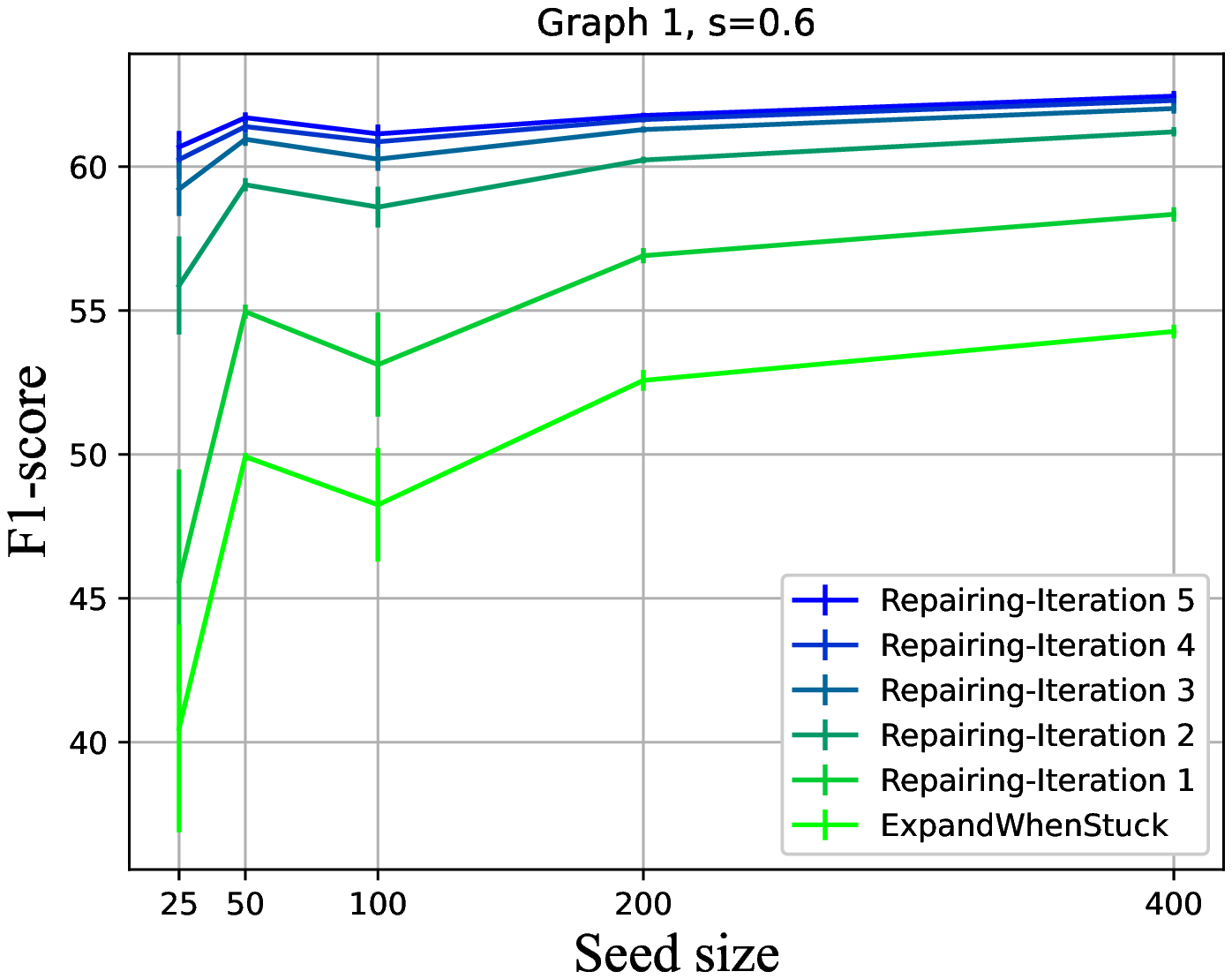}%
        }%
     \subfloat b.{%
        \includegraphics[width=0.33\linewidth]{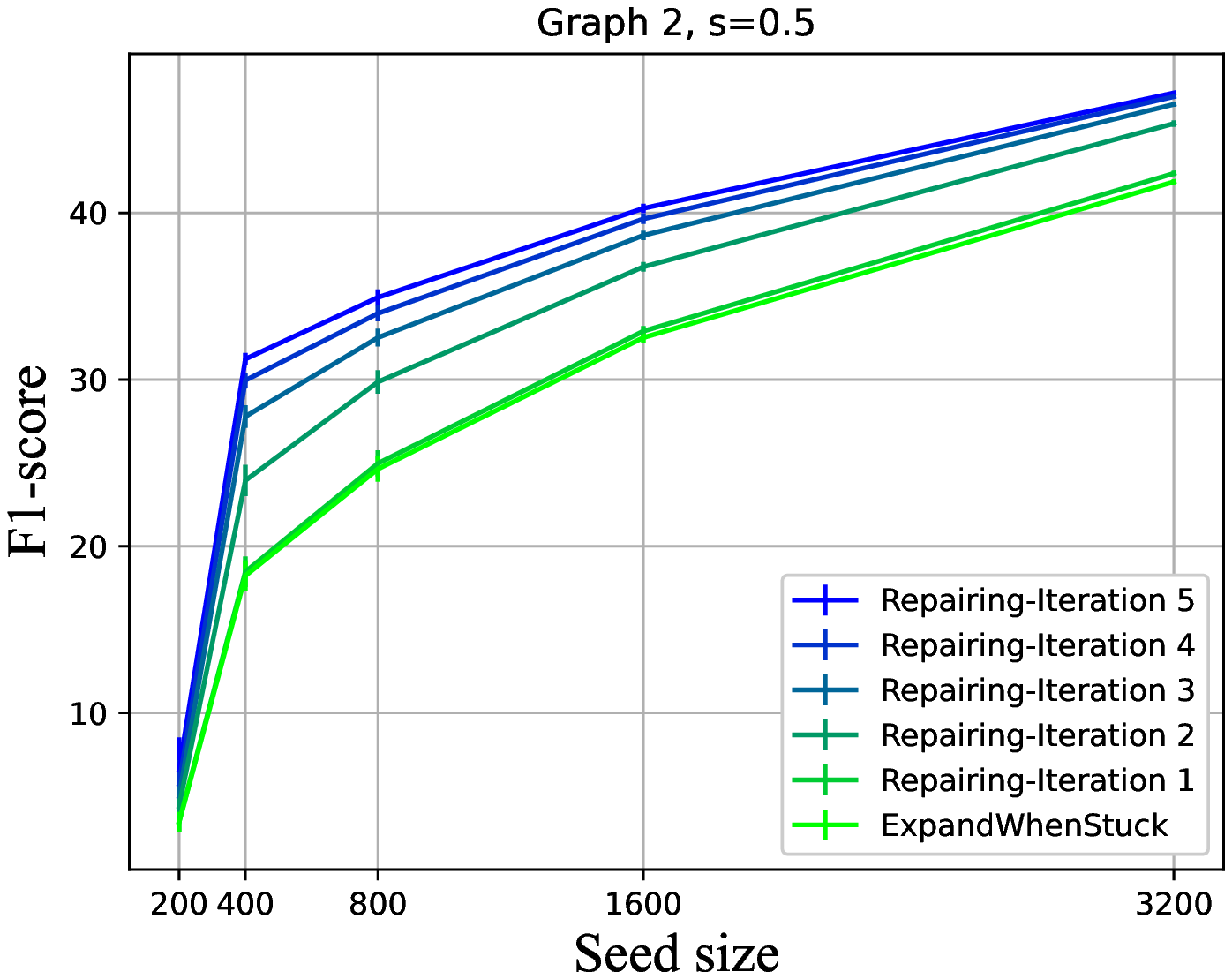}%
        }%
    \subfloat c.{%
    \includegraphics[width=0.33\linewidth]{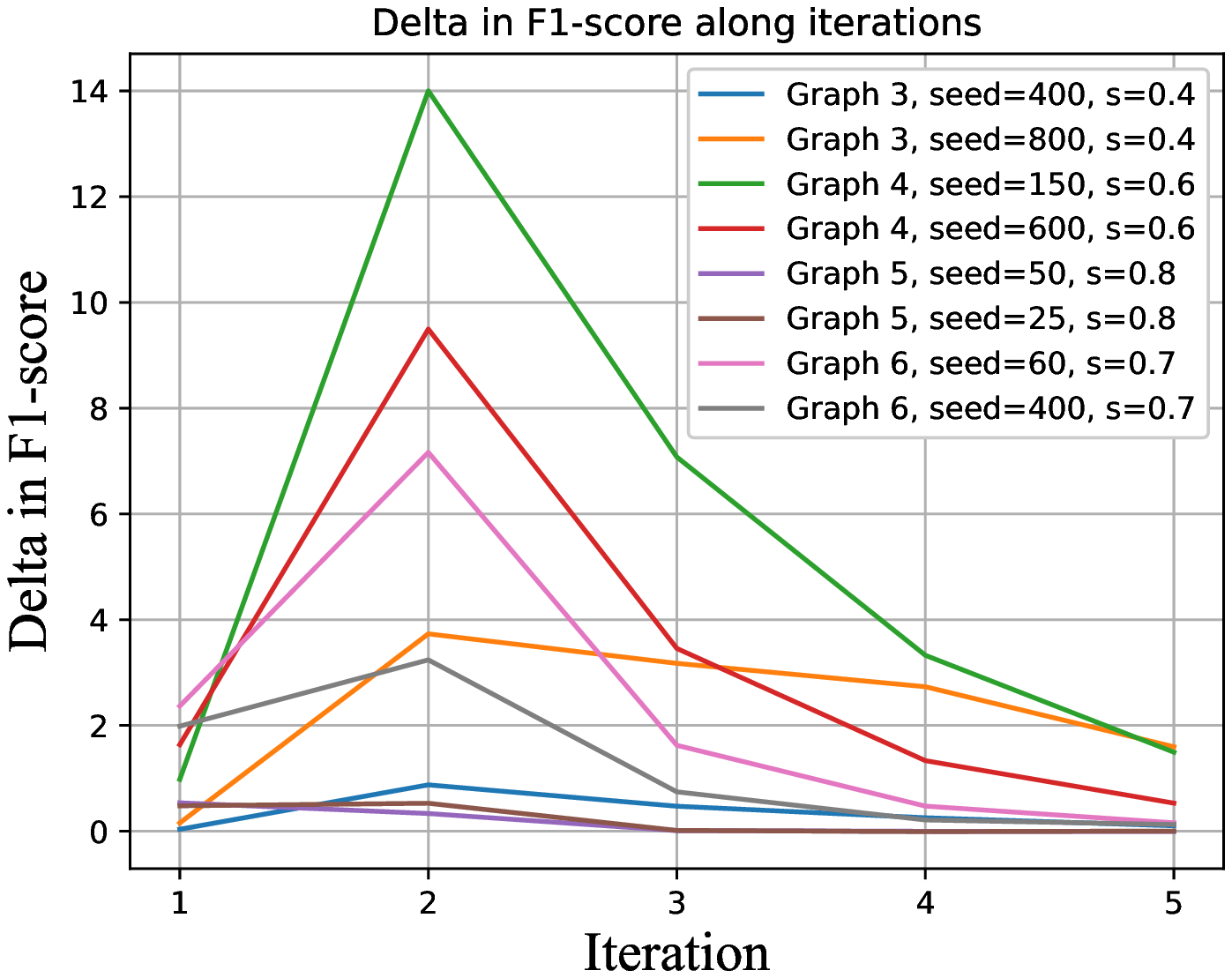}%
    }%
   \caption {Performances of IRMA: Sub-figures a,b consists of five runs of IRMA with different sizes and different seeds on a fixed graph and graphs-overlap ($s$). The value is the average over the five runs, and the error bars are the standard error.  The iterations gradually converge from EWS (Green line) to much higher values through IRMA (the blue line). The graph and $s$ value used for each sub-plot are marked on top of the plot. In sub-figure c we combined eight different runs of IRMA to present the influence of each iteration over the f1-score.}
   \label{fig:IRMA}
\end{figure*}

To test the improvement in accuracy, we computed the F1 score for all the real-world networks above (Figure \ref{fig:IRMA}). Sub-plots a and b consists of several runs of IRMA with different sizes of seeds, with a fixed graph and graphs-overlap ($s$). Each value is the average of 5 repetitions. The iterations are colored from green to blue. The lowest values are always of EWS, and the results keeps improving until they converge. Using the same experiment over other graphs, we generated sub-plot c, where each line is one IRMA run and it represents the difference in F1 (axis y) between following iterations as a function of the iteration (axis x). All positive values (as they all are) represent an improvement.

The first, most noticeable observation is that IRMA indeed gradually repairs the output map of EWS. In fact, the F1-score results are a monotonic non-decreasing function of the iterations. Moreover, the convergence is very rapid, and it  rarely takes more than 5 or 6 iterations until convergences is obtained. Finally, the most interesting thing is that IRMA is less sensitive to initial seed. EWS often achieves lower performances on a larger seeds following fluctuations in the seed composition (see sub-figure a for such a case). Despite accepting EWS's output as an initial map, IRMA exhibits a dependency on the problem properties (as the input graph and the graphs-overlap) rather than the output of EWS, and has much smoother F1 values. Importantly, the recall (and thus the F1 value) of IRMA, as presented here,  does not rise to 1. Instead, it rises to the maximal number of vertices that can have at least two marks. However, it does so even for small seeds.
\subsubsection{Stopping condition}
While the convergence of IRMA in F1 is clear from Figure \ref{fig:IRMA}, in real life scenarios, the real map is not known, and F1 cannot be used as a stopping condition. We thus propose to use  $weight(M_i)$ as a quality measure for $M_i$, and hence as a stopping condition for the algorithm. 
We recall that $weight(M)$ is defined as $|\{[u,v] |[u,v]\in G_1, [M(u),M(v)] \in G_2 \}|$ i.e.,  the number of edges in the common subgraph induced by the $M$. 
Since correct pairs share more common neighbors on average, we expect better maps to have higher $weight(M)$. Similarly,  if $weight(M_i) < (1+\delta)weight(M_{i+1})$, one can assume that $M_{i+1}$ is not significantly better than $M_i$.
We run IRMA on graphs 1 with $s=0.6$ and seed size of 50 respectively, and compared $weight(M_i)$ to the F1-score of $M_i$ to test the connection between them (Sub-figure \ref{fig:exploring_break_cond a}), other settings showed similar results. One can clearly observe that both indices are converging simultaneously and have very similar dynamics (albeit different values as expressed by the different $y$ axes in Figure Sub-figure \ref{fig:exploring_break_cond a}), suggesting that $weight(M)$ can serve as a proxy for the $F1$ score.
\begin{figure*}[h]
    \centering
    \subfloat a.{%
        \includegraphics[width=0.25\linewidth]{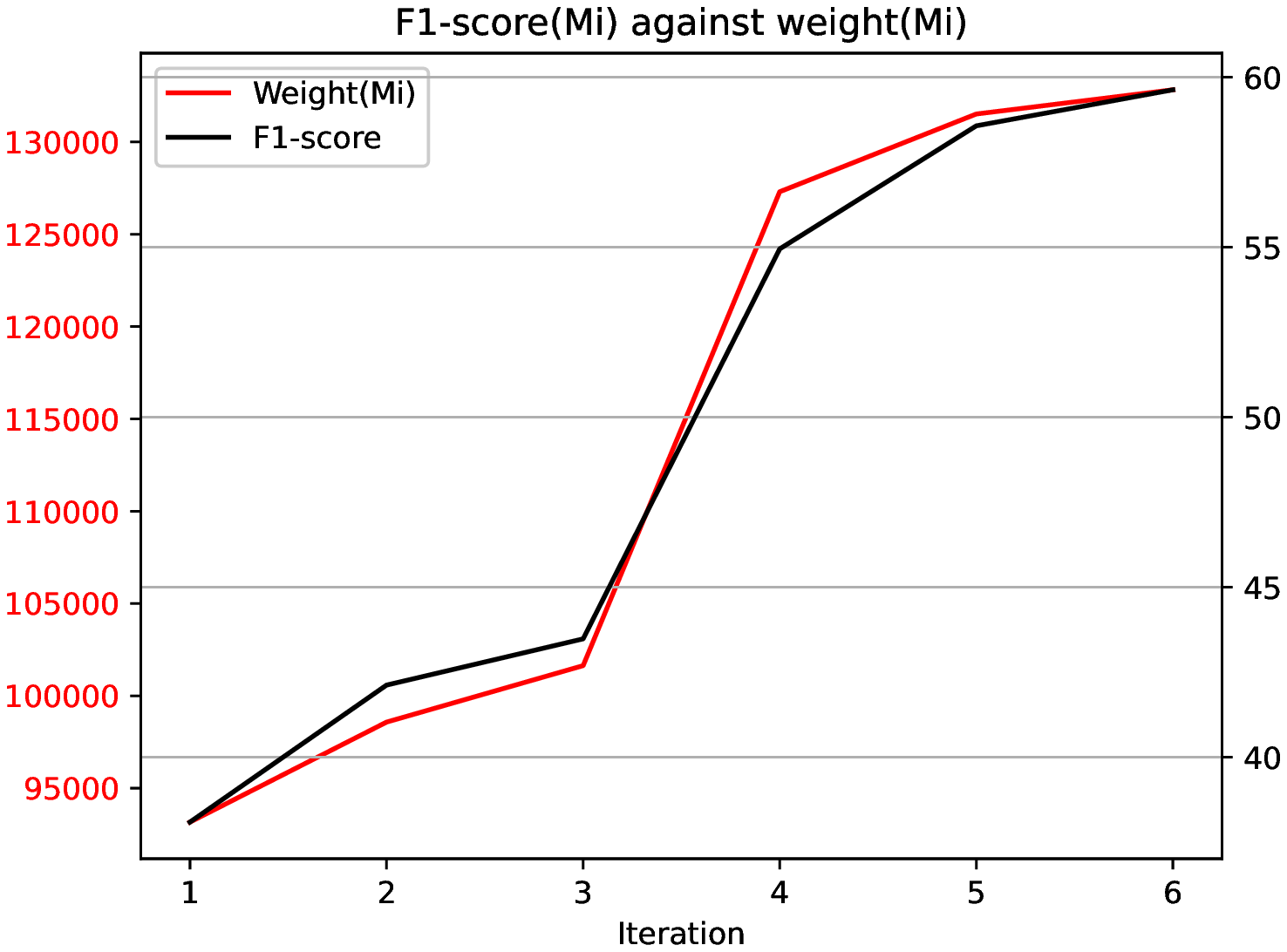}%
        }%
    \subfloat b.{%
        \includegraphics[width=0.25\linewidth]{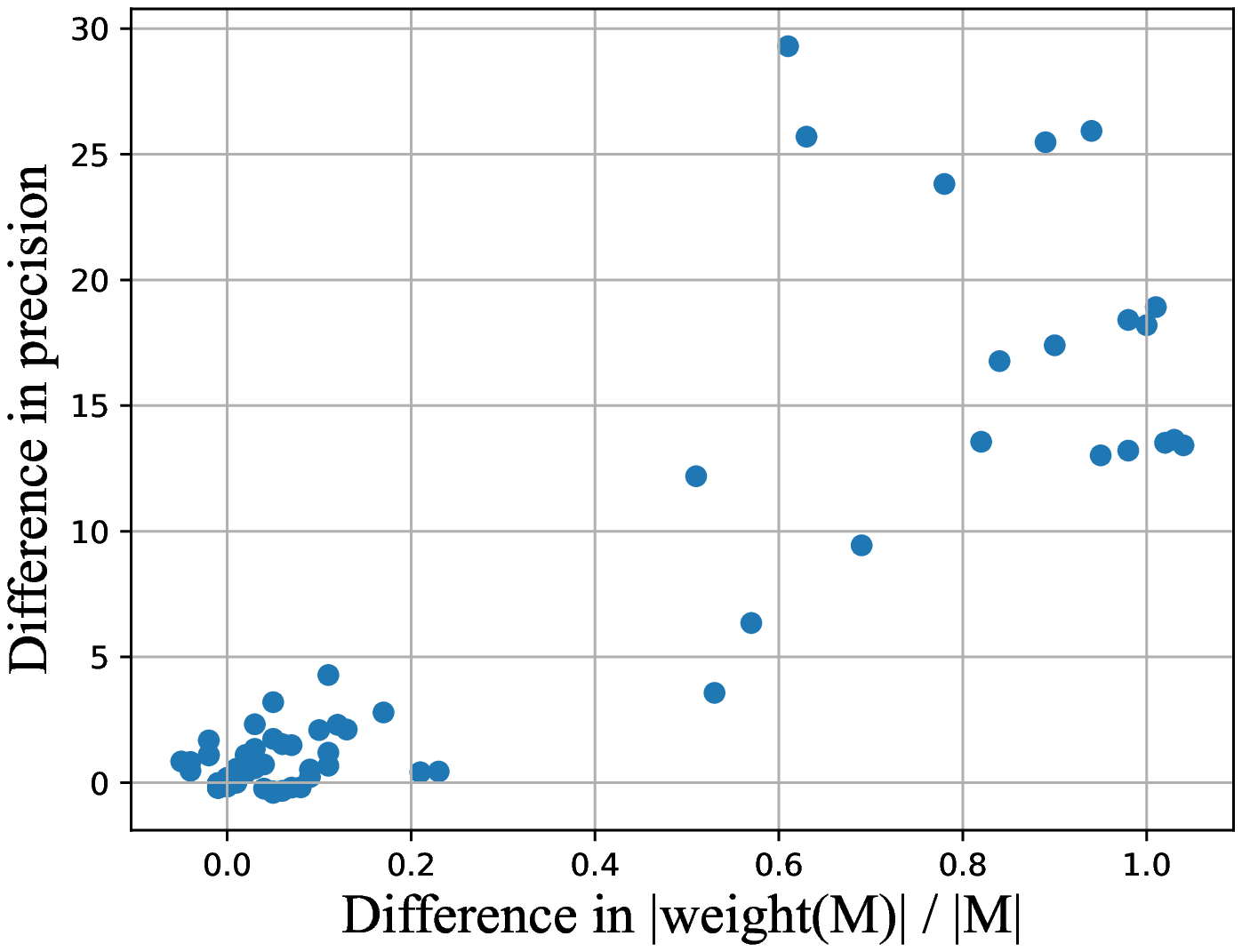}%
        }%
     \subfloat c.{%
        \includegraphics[width=0.25\linewidth]{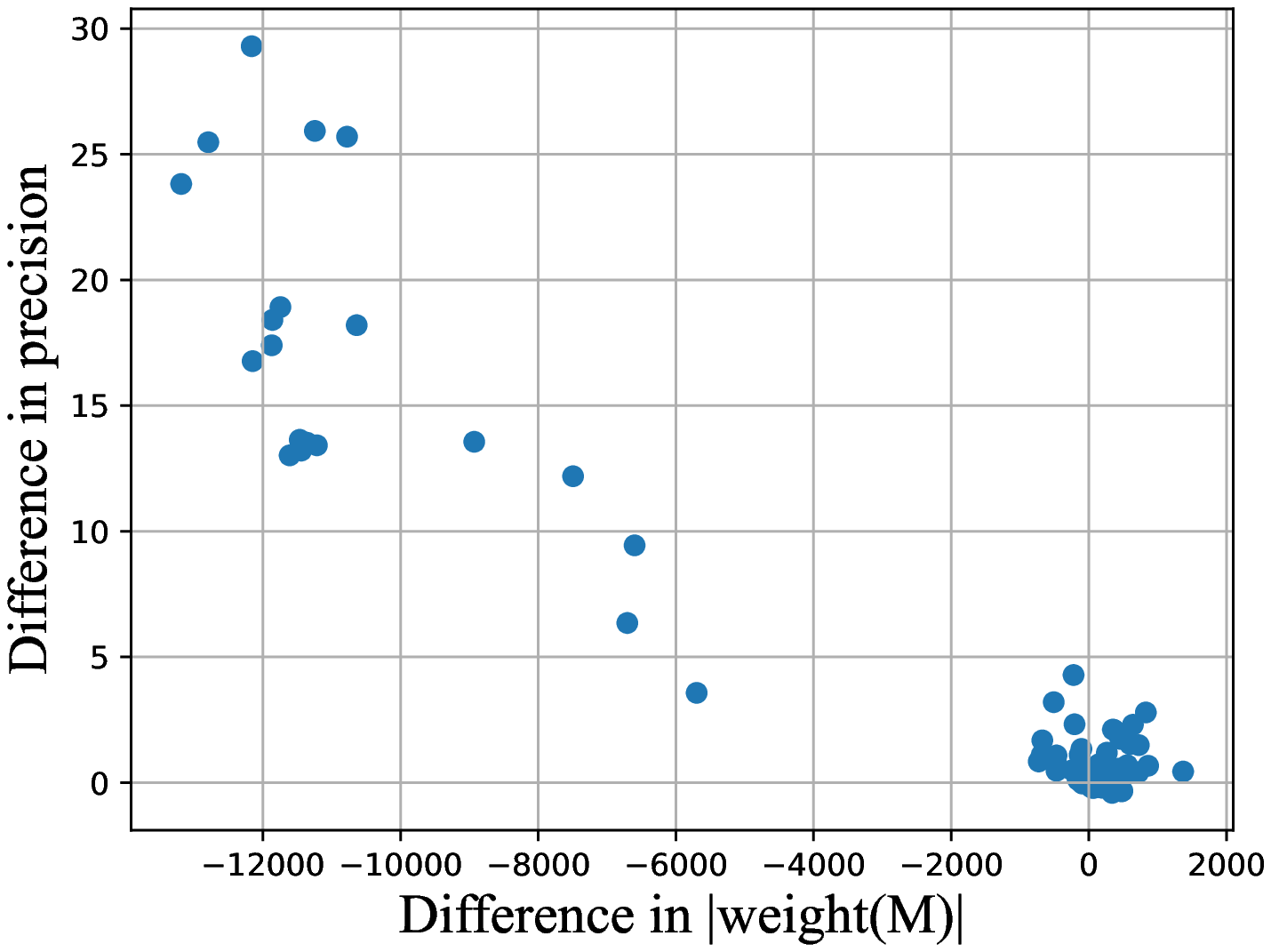}%
        }%
     \subfloat d.{%
        \includegraphics[width=0.25\linewidth]{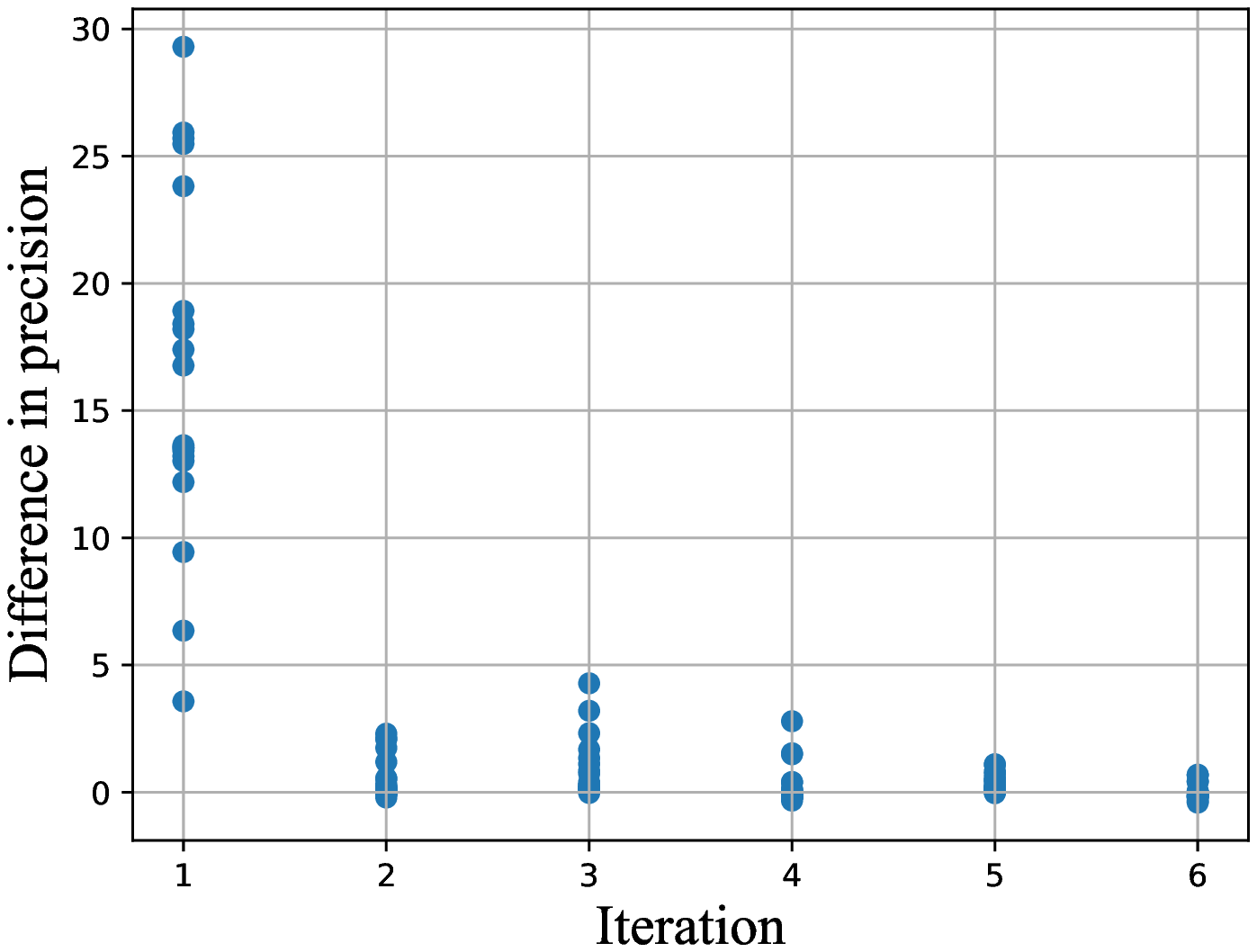}%
        }%

   \caption { 
   \textbf{a} -$weight(M)$ and F1-score of a single run of IRMA along the iterations using graphs 1 with $s=0.6$ and 50 vertices seed. \textbf{b-d} - difference in precision as a function of three indices for IRMA iterations after the exploration iteration. We used graphs 1,2,5 and 6 with seed size of 50, 400, 300 and 120 respectively and with $s=[0.5,0.6,0.7,0.8]$.  }
   \label{fig:exploring_break_cond}
\end{figure*}
\subsubsection{Computational cost of IRMA}
For a pair $[u,v] \in M$, the number of marks to spread out is $d_1(u)*d_2(v)$ where $d_i(u)$ is the degree of $u \in G_i$. Thus, the number of updates to the priority queue from inserting new pairs to $M$ is $N = \sum_{[u,v]\in M}{d_1(u)*d_2(v)}$. The cost of each insertion is proportional to the log of the size of the priority-queue, which is bound $O(|V|^2)$, so the cost of each insertion is bound by $O(Log(|V|)$, where $|V|$ is the number of vertices. The total number of updates ($N$) is between $N=O(|V|*E[d]^2)$ and $N=O(|V|*E[d^2])$, where $d$ is the degree. The first case is for random matching (i.e., $d(u)$ and $d(v)$ are independent), and the second case is for perfect matching $d(u)=d(v)$.  In low variance degree distributions, both cases are equivalent. However, in power-low degree distributions, the second case may be much higher than the first. Thus, one can bound the cost of the algorithm by $O(|V|*log(|V|)*E[d^2]$, but often the bound is tighter - $O(|E|^2*log(|V|)/|V|)$.
The Repairing-Iteration has a similar cost, since it is based on the same logic of pulling pairs from the priority queue and spreading out marks.
Note that we avoid the problematic case in EWS that the expanded seed may be of $O(|V|^2)$, which adds a $V$ factor to the cost of spreading marks.
In their parallel versions, the boundary is similar, but in practice the Repairing-Iteration takes about $30\%-60\%$ the run time of EWS, and as IRMA runs Repairing-Iteration a few times, it has an overall run time 2-3 times higher than  EWS, but of the same order. Note that the cost of IRMA as is the case for EWS is much more sensitive to the details of the propagation than to the specific number of vertices or edges. For example, IRMA has similar run times for graphs 1 and 2 used here, although graph 2 has twice as many vertices.
\subsection{Expansion to Low Degree Vertices}
The Seeded GM problem is a variant of the Maximum Common Edges Subgraph (MCES) problem, with the twist that  in seeded GM the target is to use the edge match as a tool to discover an a priori existing match and not the maximal sub-graph. Thus, in contrast with  MCES, in seeded GM, one tries to maximize not only the recall, but also the precision. Hence, at some stage, IRMA does not add low-probability pairs.

In IRMA, we use a threshold of 2 marks to add a candidate pair into $M$. Since the Repairing-Iteration method filters  wrong matched pairs, it is possible to allow the algorithm to make mistakes in some cases to find new correct pairs. We thus suggest to perform an exploration iteration with a threshold of 1 mark, after convergence, followed by regular repairing iterations to restore the precision.

Since repairing iterations filter out many wrong pairs, $weight(M_i)$ is no longer a good  condition for testing the convergence.
We tested three possible break conditions to the second stage of IRMA: A) The quality of a bijection can be defined through $Weight(M_i) / |M_i|$, where $|M_i|$ is the number of pairs in the set of matched pairs at the end of the $i$ iteration - $M_i$, and  $Weight(M_i)$ is the number of edges among members of $M_i$ shared by $G1$ and  $G2$.B)  As the target is to reduce $M$ down, we tested the condition $|M_i| \geq (1- \delta)*|M_{i-1}|$ to avoid many redundant iterations. C)  Empirically after a few iterations there is no significant improvement from $M_i$ to $M_{i+1}$. We tested a constant number of iterations.

We computed the difference in accuracy along snapshots after the exploration iteration and compared it to these  different measures (Figure \ref{fig:exploring_break_cond}). We used graphs 1,2,5 and 6 with seed size of 50, 400, 300 and 120 respectively and with all $s$ value in $[0.5,0.6,0.7,0.8]$ and fixed 6 repairing iterations. Sub-plots a and b represent the difference in precision as a function of the difference in $|weight(M_i)|/|M_i|$ and $|M_i|$, respectively. A clear correlation between the measures can be observed, yet the variance around the origin (i.e., when IRMA converges) is relatively large. Sub-plot d indicates that the most reliable prediction to the convergence of precision is the number of iteration. We use that as the stopping criteria after the exploration iteration.

\begin{figure*}[h]
    \centering
    \subfloat a.{%
        \includegraphics[width=0.3\linewidth]{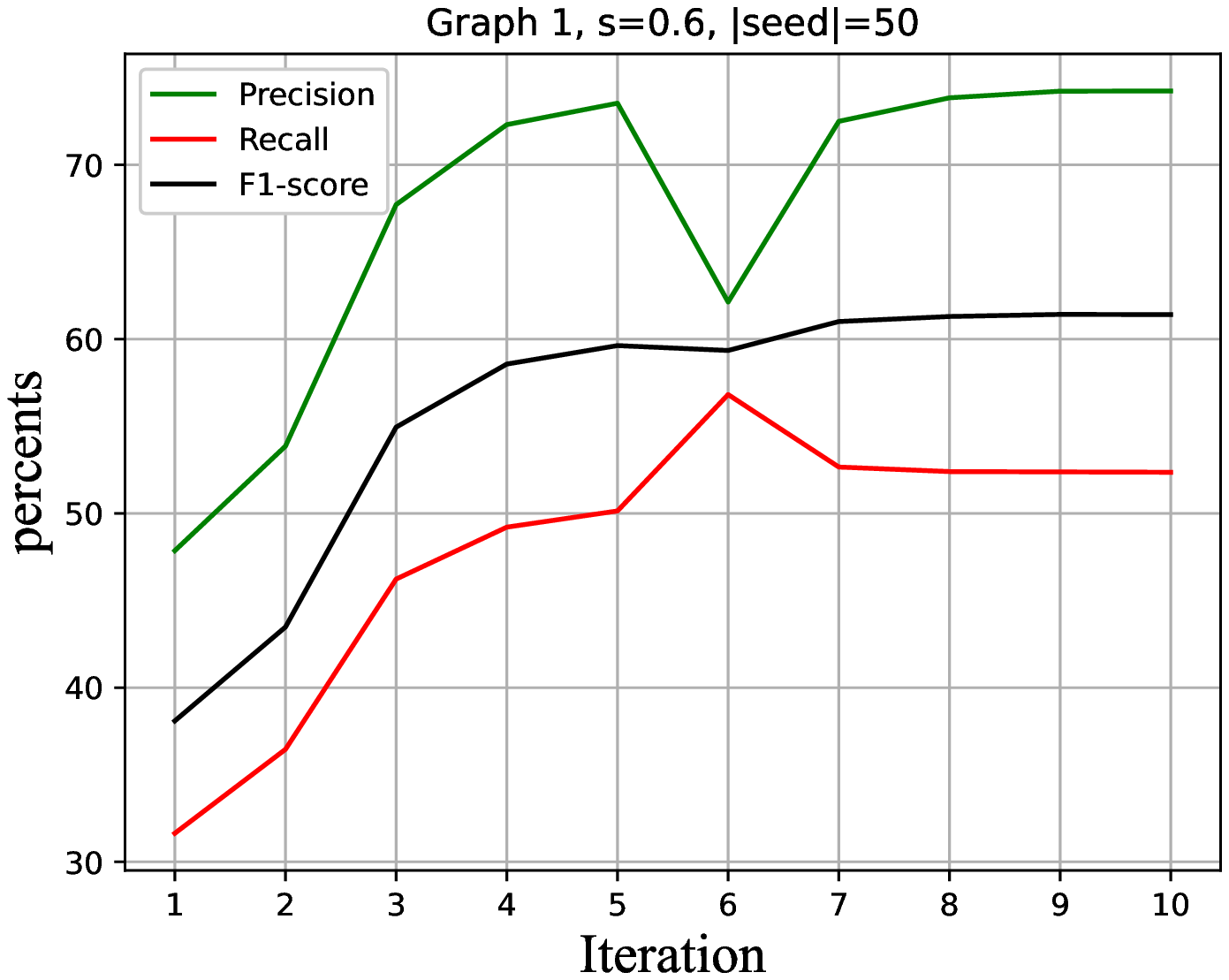}%
        }%
     \subfloat b.{%
        \includegraphics[width=0.3\linewidth]{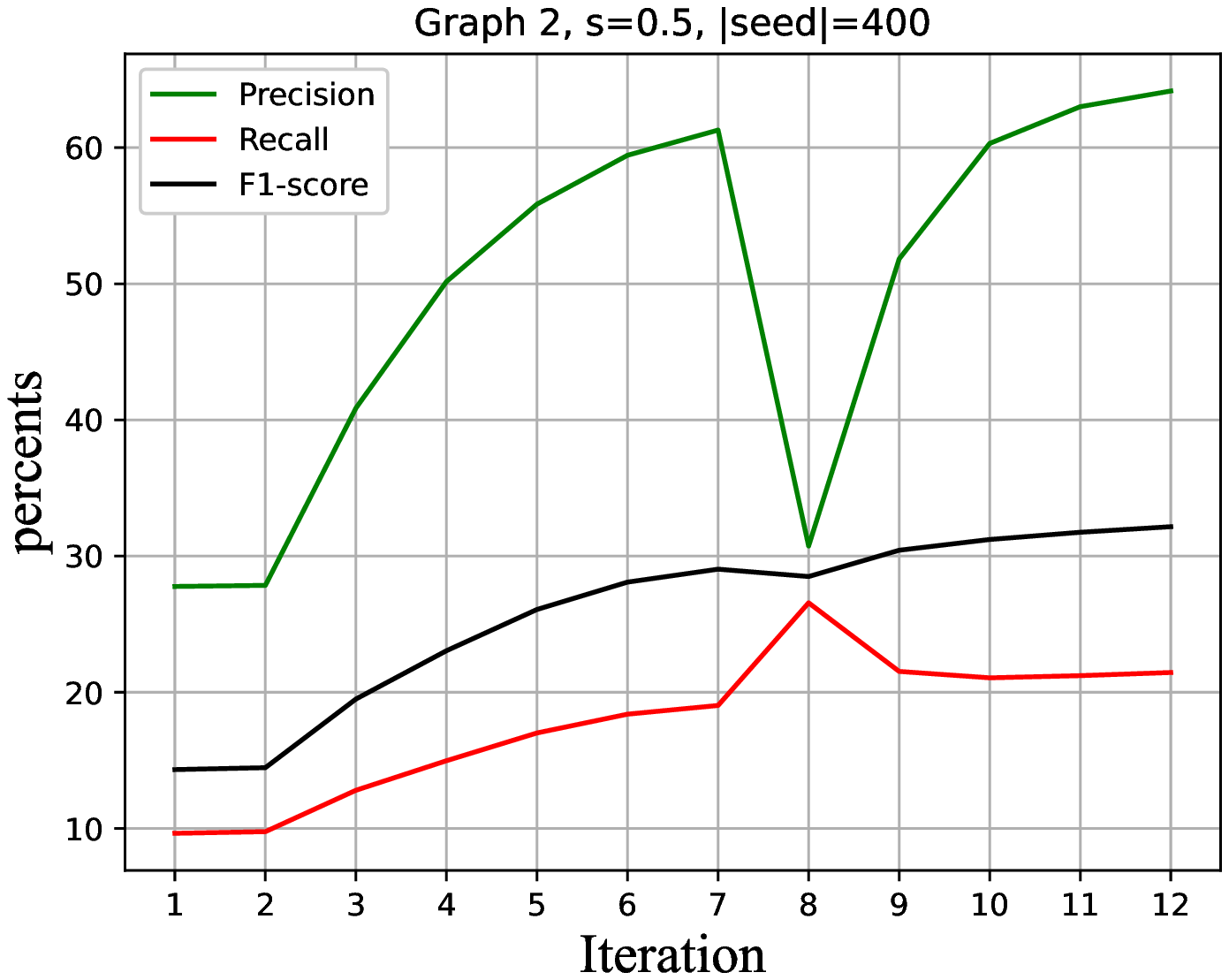}%
        }%
     \subfloat c.{%
        \includegraphics[width=0.3\linewidth]{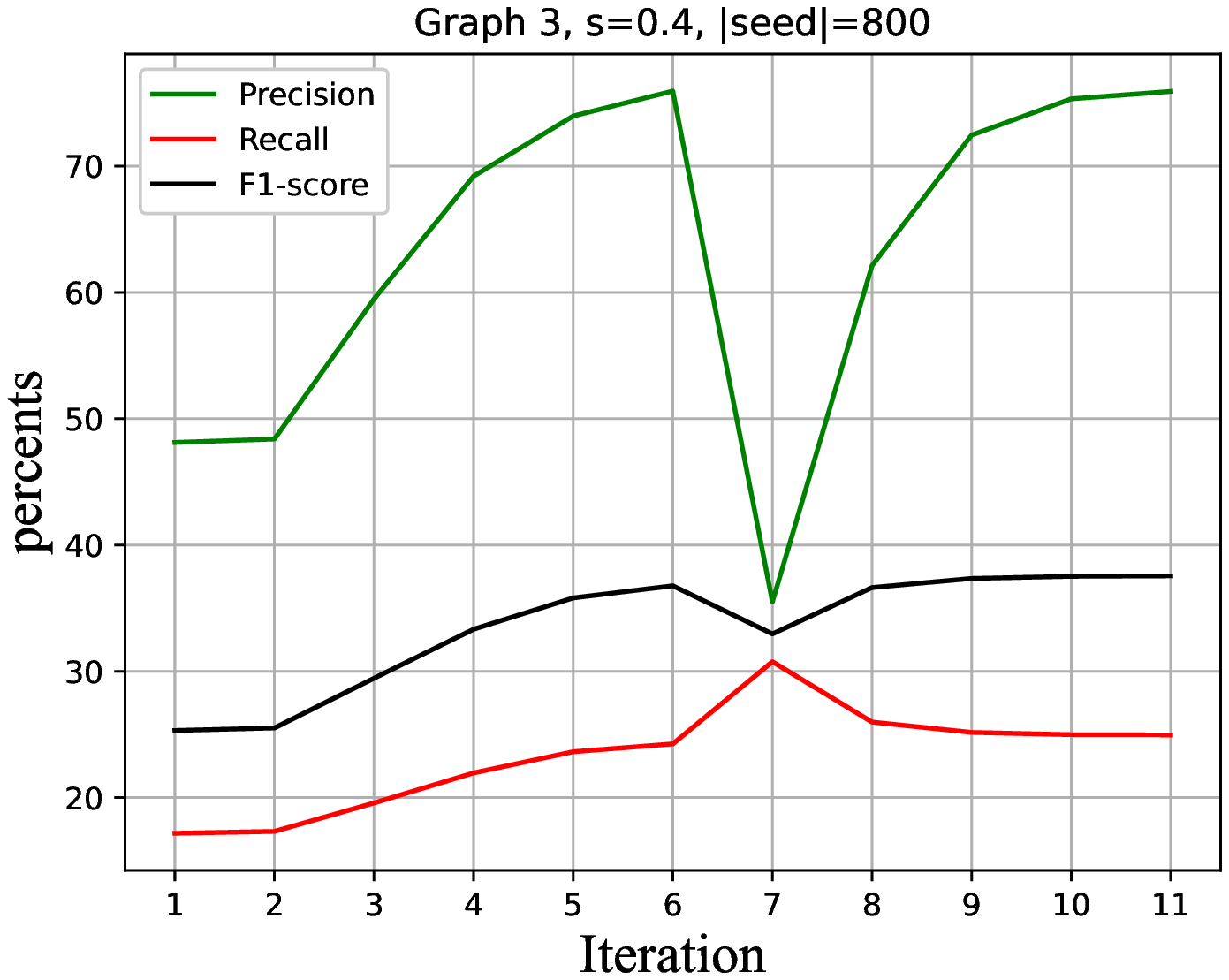}%
        }%
 \caption {Precision, recall and F1-score during the algorithm's iterations: Each sub-figure consists of a single run of IRMA with the exploring iteration. We used graphs 1-3 with seed size of 50, 400, and 800 respectively and with $s$ value of $0.6, 0.5$ and $0.4$ respectively. The drop in precision is caused by the exploring iteration, after which the precision is fully restore while recall stays higher than before.}
    \label{fig:precision_and_recall}
\end{figure*}

We used these additional standard IRMA iterations after the exploration iteration, and tested the precision, recall and F1-score along IRMA's iterations on graphs 1-3 with seed size of 50, 400 and 800 respectively and with $s$ value of $0.6, 0.5$ and $0.4$ respectively (Figure \ref{fig:precision_and_recall}). The relation between the precision and the recall is the ratio  between $R$ and $M$. Both recall and precision are monotonic non decreasing up to the exploration iteration, followed by a drop in precision together with the increase in recall in the exploration iteration, and the restoration of the  precision in 3-4 iterations. Note that while precision is fully restored (and some times even improves), the recall stays higher than before the exploration iteration. In theory, one could repeat the process of doing exploration and repair iterations. However, in practice, the $F1$ gain of multiple cycles is very small.

\begin{figure*}[]
    \centering
    \subfloat a.{%
        \includegraphics[width=0.3\linewidth]{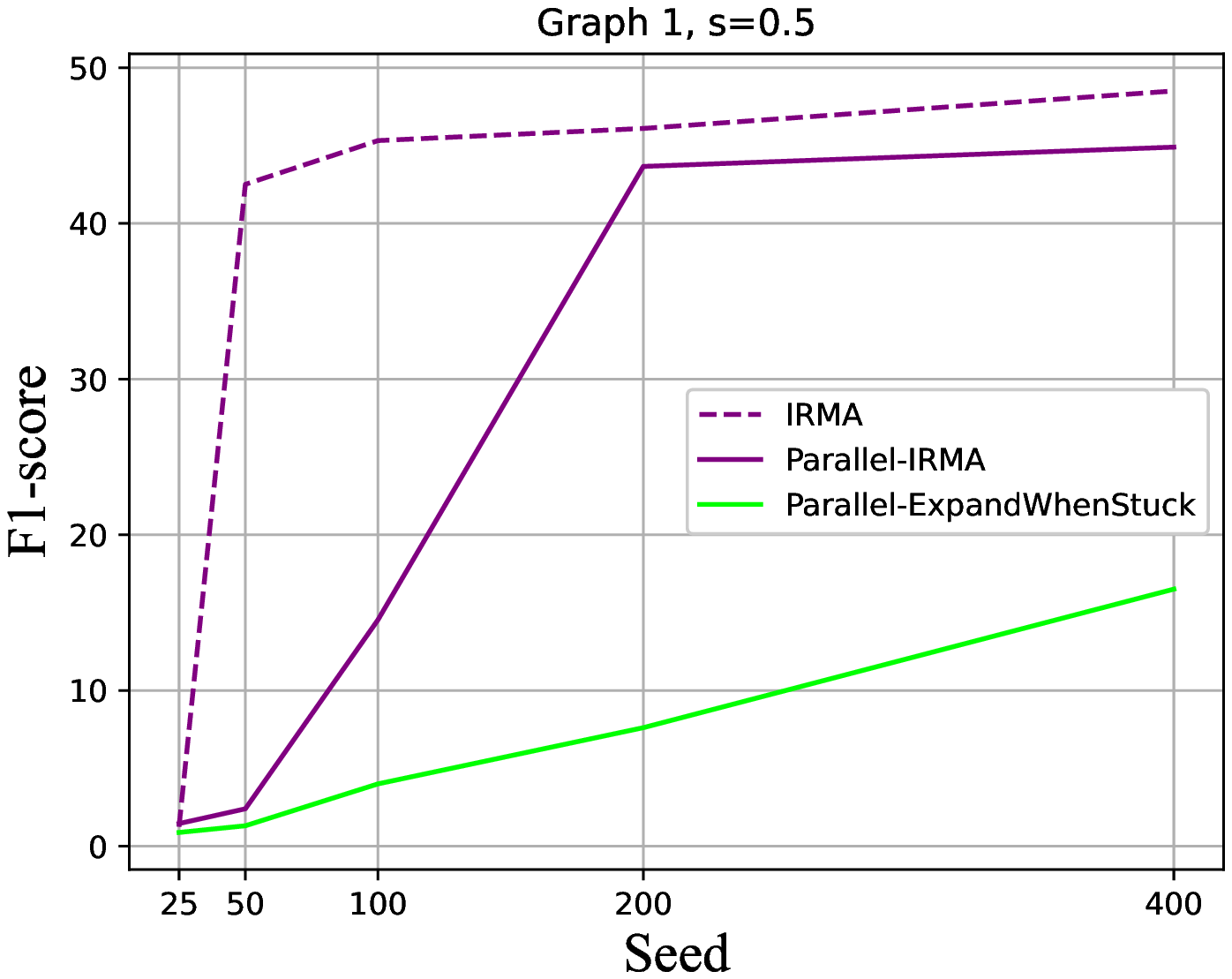}%
        }%
     \subfloat b.{%
        \includegraphics[width=0.3\linewidth]{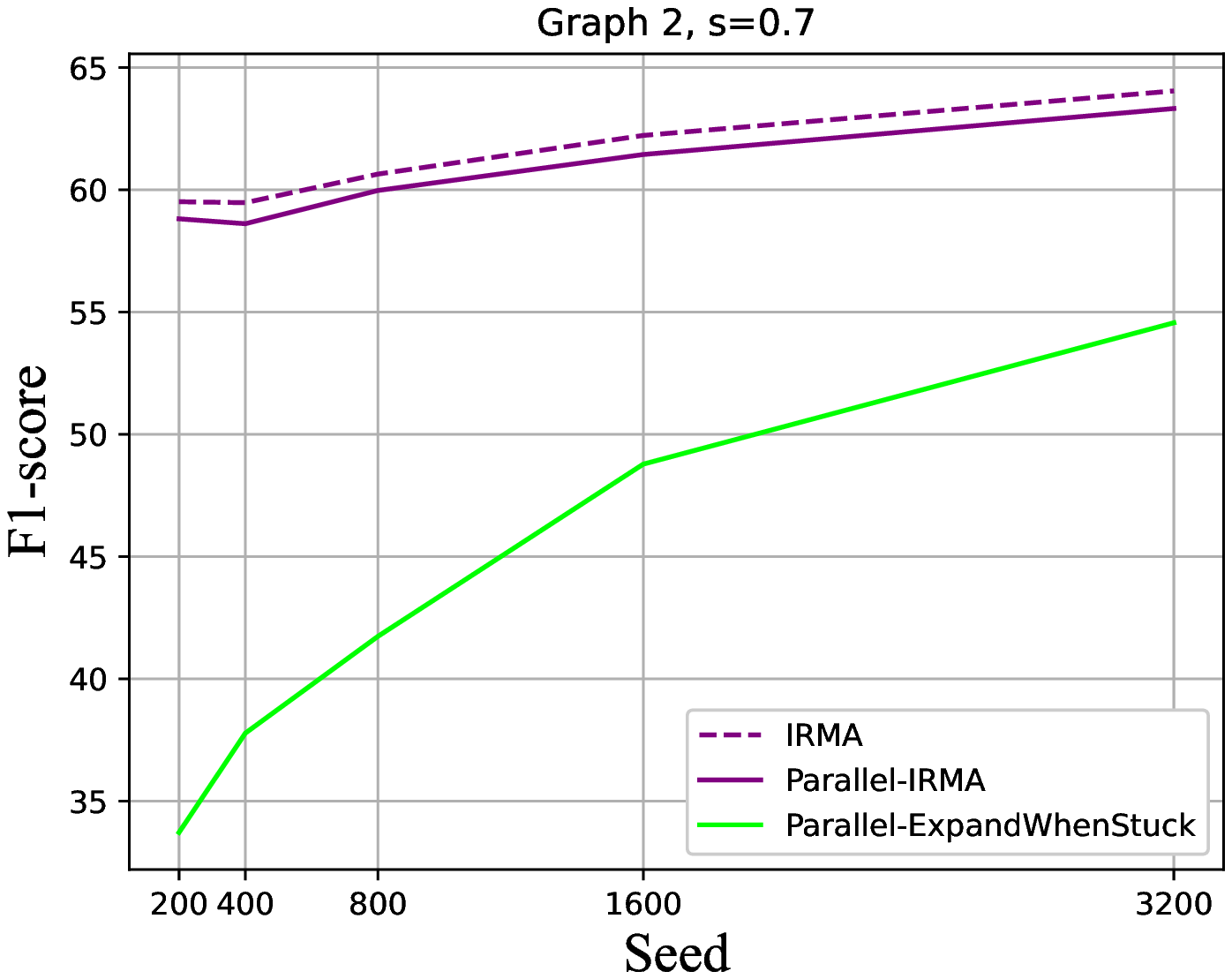}%
        }%
     \subfloat c.{%
        \includegraphics[width=0.3\linewidth]{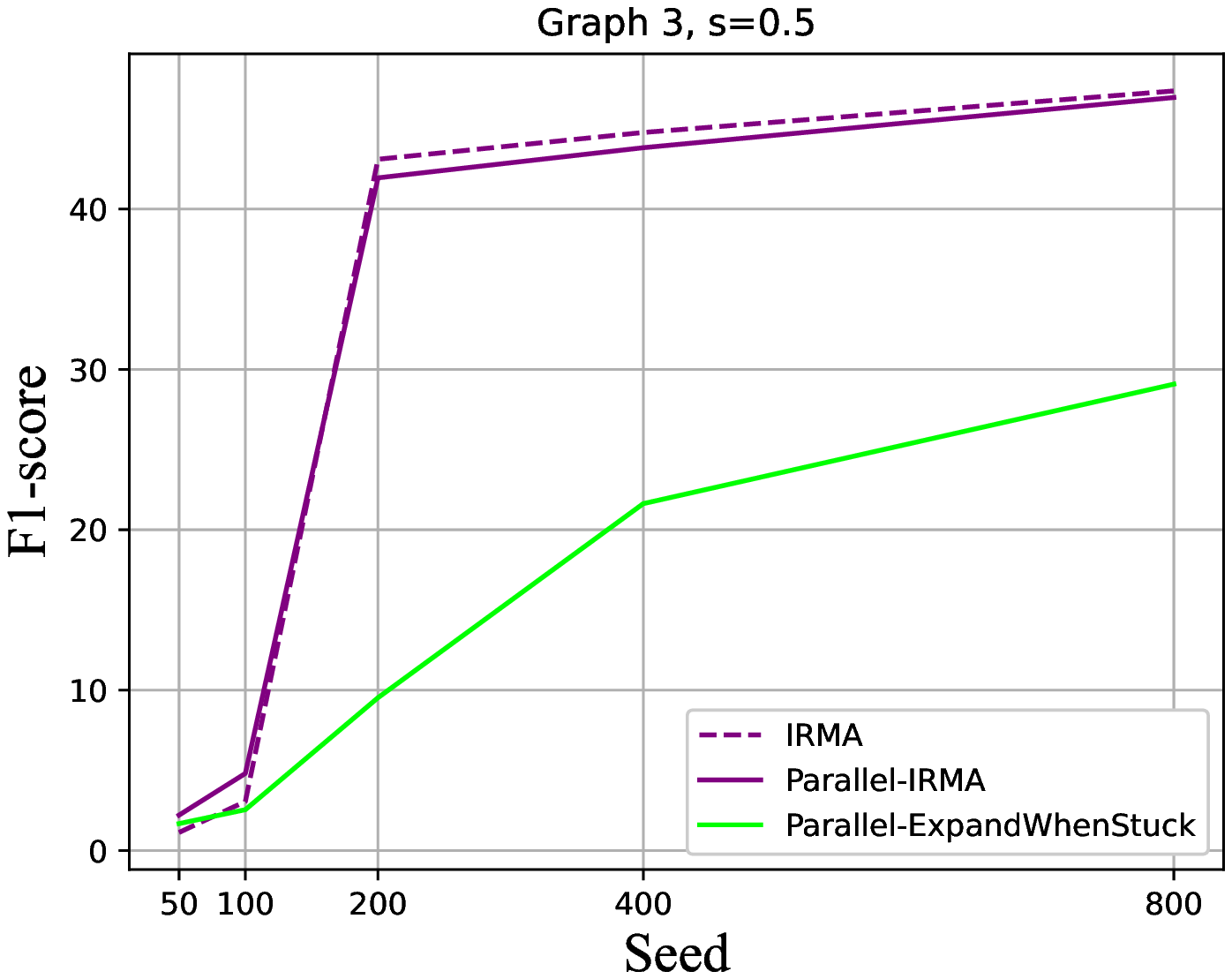}%
        }%
   \caption {A comparison between parallel-ExpandWhenStuck, Parallel-IRMA and regular IRMA (both use an exploring iteration).}
   \label{fig:parallel_main_plots}
\end{figure*}
\subsection{Comparison to existing methods}
Current state of the art seeded GM for scale free graphs include beyond EWS two main methods. Yu et al \cite{yu2021power} proposed The D-hop method based on diffusing the seed more than one neighbor away (denoted as PLD). Chiasserini et al \cite{chiasserini2016social} published the DDM de-anonymization algorithm (which is  equivalent to seed GM) based on parallel graph partitioning. The two studies report different measures on different graphs. 

\begin{figure}
    \centering
        \includegraphics[width=0.5\linewidth]{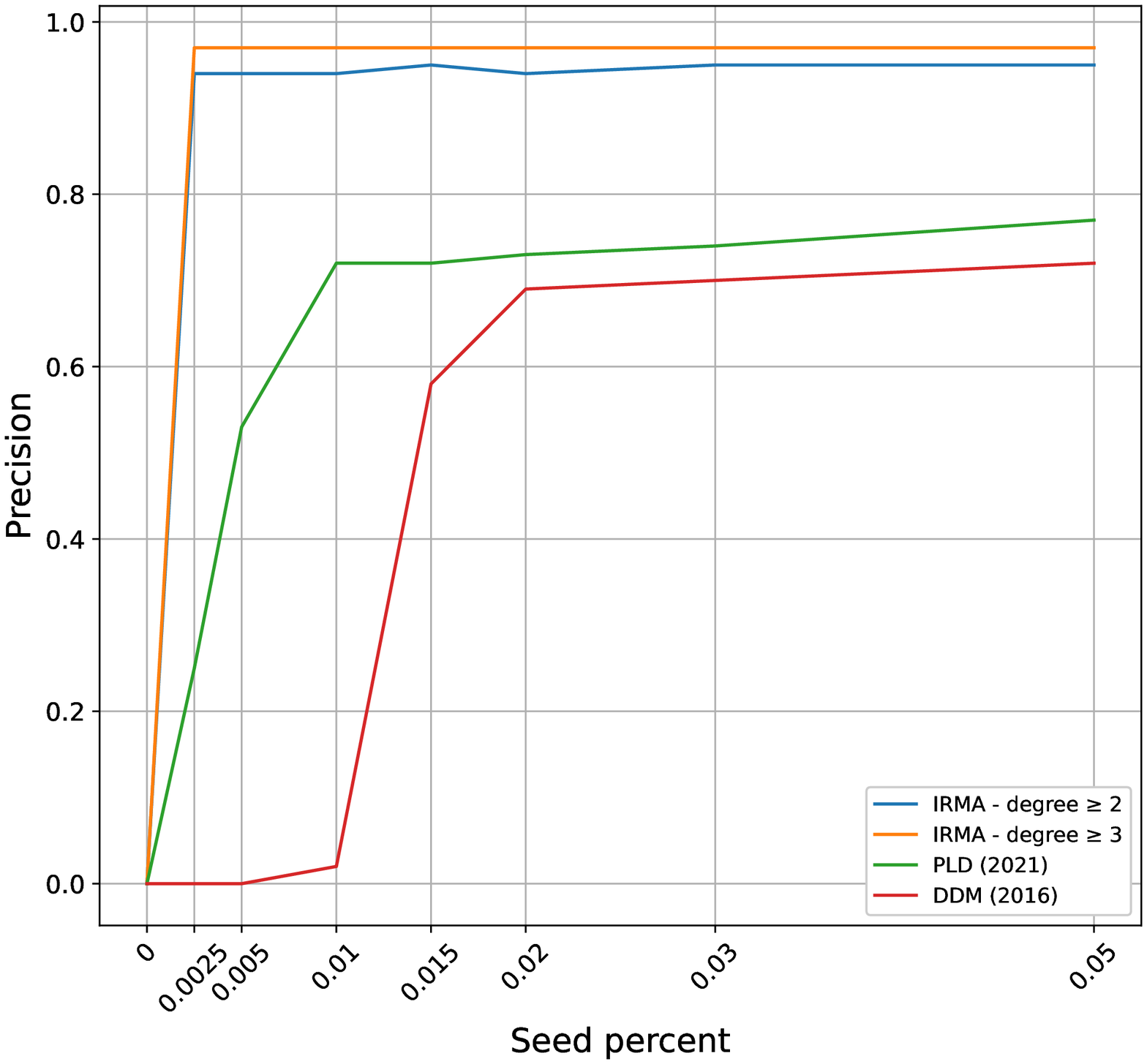}%
        \includegraphics[width=0.5\linewidth]{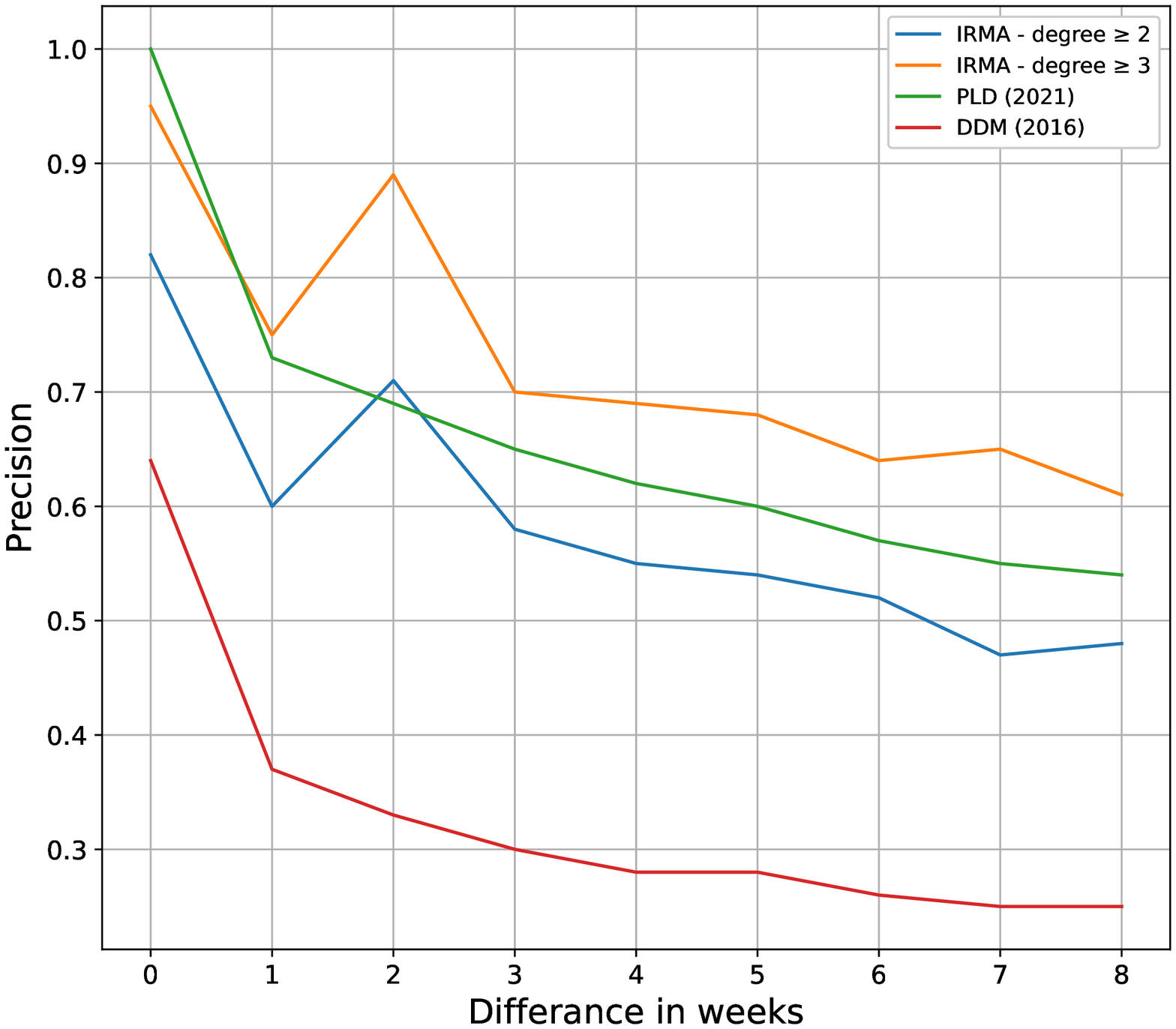}%

   \caption {reproduced the experiment from PLD and DDM. sub-figure a used the Facebook data-set using sampling with s=$0.9$, t=$0.8$ and different seed size. In sub-figure b we used AS data-set with seed size = $0.01\%$. PLD uses a cutoff on the degree for the analysis and IRMA uses even lower degree vertices and is thus more inclusive}
    \label{fig:comparison_to_existing}
\end{figure}

To show that IRMA obtains better results than both algorithms, we repeated their analysis using the setting used in each algorithm, and the accuracy measure. The analysis was performed on the FaceBook and AS datasets (Figure \ref{fig:comparison_to_existing}). PLD uses a cutoff on the vertices degree that depends on the details of the algorithm, but in their setting is higher than 3. WE thus show our results on vertices of degree above 2 or above 3. In all cases, the precision of IRMA is higher on the same set of verices.

\subsection{Parallel Solution}
In the parallel version to EWS. The main idea is to split the algorithm into epochs such that marks are spread out only between the epochs, allowing parallel spreading. Using the same concept, we  developed  a parallel version to Repairing-Iteration in which $M$ is rebuilt based only on the marks of the former iteration. The parallel version of IRMA  starts by running Parallel-EWS and then performs iterations of Parallel-Repairing-Iteration (see section \ref{sec:Iterative_approach}).

When comparing (Figure. \ref{fig:parallel_main_plots}) the parallel-EWS, Parallel-IRMA and regular IRMA (both use exploring iteration). The parallel-IRMA has much better precision and recall than the Parallel-EWS. In sub-plot E, one can see an extreme case where Parallel-IRMA with a seed of 25 pairs is more accurate than Parallel-EWS with 400 pairs in the seed. The Parallel-IRMA and the standard IRMA have very similar recall and precision, making the parallel IRMA an excellent trade-off between run-time and performances. Note that the parallel IRMA has the same run time as the parallel EWS up to multiplication by a small constant, and is much faster, yet much more accurate than the non-parallel EWS.

\section{Discussion}
We have here extended the EWS seeded-GM algorithm to include iterative corrections of the matching. EWS spreads of marks to common neighbors of previously matched pairs. It is greedy in the sense that once a pair has been matched, it cannot be changed. 

We have here shown that real matches tend to accumulate more marks than wrong ones, even when a wrong match has been inserted to the match before the real one, preventing it from being added to the match. As such, one can repair at the end of each iteration the paired matches, and repeat the iteration. The resulting iterative algorithm is named IRMA, which has a better accuracy than the current state-of-the-art GM algorithms.

While in $G(n,p)$ graphs existing seeded GM algorithms get a very high accuracy even for a small seed, the same does not hold for scale-free networks, as are most real-world networks. There are two main reasons for the failure of seeded GM in scale free networks: A) pairs of very high degree vertices (that are not real pairs) easily accumulate wrong marks. In a greedy approach, high-degree vertices will receive a lot of marks from common neighbors, since they tend to have a lot of random common neighbors. B) Very low degree vertices have few neighbors, and as such receive very few marks, and may never be explored. 

IRMA solves the first problem by iteratively fixing errors. However, a simple iterative correction cannot solve the second problem. To address the second problem,  an expansion iteration was added, where IRMA accumulates a large number of low-degree pairs with a large fraction of wrong-pairs. This is then followed by multiple correction iterations to improve the accuracy, while maintaining a high recall. 

Each iteration of IRMA is a full iteration of EWS, and typical applications require 5-10 iterations. The run time could be expected to be 5-10 time longer than EWS. However, in practice, the gain in accuracy of IRMA is much higher than the accuracy difference between the parallel and iterative versions of EWS. As such a parallel version of  IRMA can be used which is faster than regular EWS with higher precision and recall.

We have tested  alternative methods to fix wrong pairs in each iteration, including the analysis of negative marks (marks from contradicting neighbors), or the relation between marks and degree. However, eventually, the simplest approach to accumulate marks, even after a pair was selected, ended up being the one best improving the precision and recall.

The extension of seeded GM algorithms to scale-free network with limited overlap (low values of $s$) is essential for their application in real-world networks. However, the model we have used here for the generation of partially overlapping graphs (random sampling from a larger common graph), may not represent the real difference between networks. Thus, an important extension of the current method would be to test it on real world partially overlapping networks models (i.e two different networks, with parts of the vertices representing the same entity). 

Another important caveat of IRMA is that in the current version it is applied to unweighted undirected graphs. The extension to directed and weighted graphs is straightforward. However, this would add multiple free parameters to the analysis (such as the difference between the in and out degree distributions, and the weight distribution). We have thus discussed here the simpler case. We expect the results here to hold for more general models, and plan to further test it.

\clearpage
\bibliographystyle{plain}
\bibliography{Main.bib}

\begin{thebibliography}{10}

\bibitem{chiasserini2016social}
Carla-Fabiana Chiasserini, Michele Garetto, and Emilio Leonardi.
\newblock Social network de-anonymization under scale-free user relations.
\newblock {\em IEEE/ACM Transactions on Networking}, 24(6):3756--3769, 2016.

\bibitem{cullina2016improved}
Daniel Cullina and Negar Kiyavash.
\newblock Improved achievability and converse bounds for erdos-renyi graph
  matching.
\newblock {\em ACM SIGMETRICS Performance Evaluation Review}, 44(1):63--72,
  2016.

\bibitem{cullina2017exact}
Daniel Cullina and Negar Kiyavash.
\newblock Exact alignment recovery for correlated erd$\backslash$h $\{$o$\}$
  sr$\backslash$'enyi graphs.
\newblock {\em arXiv preprint arXiv:1711.06783}, 2017.

\bibitem{ding2021efficient}
Jian Ding, Zongming Ma, Yihong Wu, and Jiaming Xu.
\newblock Efficient random graph matching via degree profiles.
\newblock {\em Probability Theory and Related Fields}, 179(1):29--115, 2021.

\bibitem{egozi2012probabilistic(7)}
Amir Egozi, Yosi Keller, and Hugo Guterman.
\newblock A probabilistic approach to spectral graph matching.
\newblock {\em IEEE Transactions on Pattern Analysis and Machine Intelligence},
  35(1):18--27, 2012.

\bibitem{Erdos-Renyi}
P~Erd\"os and A~R\'enyi.
\newblock On random graphs i.
\newblock {\em Publicationes Mathematicae Debrecen}, 6:290--297, 1959.

\bibitem{fishkind2012seeded}
Donniell~E Fishkind, Sancar Adali, Heather~G Patsolic, Lingyao Meng, Digvijay
  Singh, Vince Lyzinski, and Carey~E Priebe.
\newblock Seeded graph matching.
\newblock {\em arXiv preprint arXiv:1209.0367}, 2012.

\bibitem{fishkind2019seeded}
Donniell~E Fishkind, Sancar Adali, Heather~G Patsolic, Lingyao Meng, Digvijay
  Singh, Vince Lyzinski, and Carey~E Priebe.
\newblock Seeded graph matching.
\newblock {\em Pattern Recognition}, 87:203--215, 2019.

\bibitem{goga2013exploiting}
Oana Goga, Howard Lei, Sree Hari~Krishnan Parthasarathi, Gerald Friedland,
  Robin Sommer, and Renata Teixeira.
\newblock Exploiting innocuous activity for correlating users across sites.
\newblock In {\em Proceedings of the 22nd International Conference on World
  Wide Web}, pages 447--458, 2013.

\bibitem{henderson2011s(10)}
Keith Henderson, Brian Gallagher, Lei Li, Leman Akoglu, Tina Eliassi-Rad,
  Hanghang Tong, and Christos Faloutsos.
\newblock It's who you know: graph mining using recursive structural features.
\newblock In {\em Proceedings of the 17th ACM SIGKDD International Conference
  on Knowledge Discovery and Data Mining}, pages 663--671, 2011.

\bibitem{EXPAND}
Ehsan Kazemi, S.~Hamed Hassani, and Matthias Grossglauser.
\newblock Growing a graph matching from a handful of seeds.
\newblock {\em Proceedings of the VLDB Endowment}, 8(10):1010--1021, 2015.

\bibitem{kazemi2015can}
Ehsan Kazemi, Lyudmila Yartseva, and Matthias Grossglauser.
\newblock When can two unlabeled networks be aligned under partial overlap?
\newblock In {\em 2015 53rd Annual Allerton Conference on Communication,
  Control, and Computing (Allerton)}, pages 33--42. IEEE, 2015.

\bibitem{klau2009new(13)}
Gunnar~W Klau.
\newblock A new graph-based method for pairwise global network alignment.
\newblock {\em BMC Bioinformatics}, 10(1):1--9, 2009.

\bibitem{korula2013efficient(14)}
Nitish Korula and Silvio Lattanzi.
\newblock An efficient reconciliation algorithm for social networks.
\newblock {\em arXiv preprint arXiv:1307.1690}, 2013.

\bibitem{koutra2013big}
Danai Koutra, Hanghang Tong, and David Lubensky.
\newblock Big-align: Fast bipartite graph alignment.
\newblock In {\em 2013 IEEE 13th International Conference on Data Mining},
  pages 389--398. IEEE, 2013.

\bibitem{malhotra2012studying(16)}
Anshu Malhotra, Luam Totti, Wagner Meira~Jr, Ponnurangam Kumaraguru, and
  Virgilio Almeida.
\newblock Studying user footprints in different online social networks.
\newblock In {\em 2012 IEEE/ACM International Conference on Advances in Social
  Networks Analysis and Mining}, pages 1065--1070. IEEE, 2012.

\bibitem{mossel2020seeded}
Elchanan Mossel and Jiaming Xu.
\newblock Seeded graph matching via large neighborhood statistics.
\newblock {\em Random Structures \& Algorithms}, 57(3):570--611, 2020.

\bibitem{narayanan2012feasibility}
Arvind Narayanan, Hristo Paskov, Neil~Zhenqiang Gong, John Bethencourt, Emil
  Stefanov, Eui Chul~Richard Shin, and Dawn Song.
\newblock On the feasibility of internet-scale author identification.
\newblock In {\em 2012 IEEE Symposium on Security and Privacy}, pages 300--314.
  IEEE, 2012.

\bibitem{nunes2012resolving(20)}
Andr{\'e} Nunes, P{\'a}vel Calado, and Bruno Martins.
\newblock Resolving user identities over social networks through supervised
  learning and rich similarity features.
\newblock In {\em Proceedings of the 27th Annual ACM Symposium on Applied
  Computing}, pages 728--729, 2012.

\bibitem{pedarsani2011privacy}
Pedram Pedarsani and Matthias Grossglauser.
\newblock On the privacy of anonymized networks.
\newblock In {\em Proceedings of the 17th ACM SIGKDD International Conference
  on Knowledge Discovery and Data Mining}, pages 1235--1243, 2011.

\bibitem{riederer2016linking}
Christopher Riederer, Yunsung Kim, Augustin Chaintreau, Nitish Korula, and
  Silvio Lattanzi.
\newblock Linking users across domains with location data: Theory and
  validation.
\newblock In {\em Proceedings of the 25th International Conference on World
  Wide Web}, pages 707--719, 2016.

\bibitem{shirani2017seeded}
Farhad Shirani, Siddharth Garg, and Elza Erkip.
\newblock Seeded graph matching: Efficient algorithms and theoretical
  guarantees.
\newblock In {\em 2017 51st Asilomar Conference on Signals, Systems, and
  Computers}, pages 253--257. IEEE, 2017.

\bibitem{singh2008global(25)}
Rohit Singh, Jinbo Xu, and Bonnie Berger.
\newblock Global alignment of multiple protein interaction networks with
  application to functional orthology detection.
\newblock {\em Proceedings of the National Academy of Sciences},
  105(35):12763--12768, 2008.

\bibitem{tan2014mapping}
Shulong Tan, Ziyu Guan, Deng Cai, Xuzhen Qin, Jiajun Bu, and Chun Chen.
\newblock Mapping users across networks by manifold alignment on hypergraph.
\newblock In {\em Twenty-Eighth AAAI Conference on Artificial Intelligence},
  2014.

\bibitem{torresani2008feature(26)}
Lorenzo Torresani, Vladimir Kolmogorov, and Carsten Rother.
\newblock Feature correspondence via graph matching: Models and global
  optimization.
\newblock In {\em European Conference on Computer Vision}, pages 596--609.
  Springer, 2008.

\bibitem{wiskott1997face(28)}
Laurenz Wiskott, Norbert Kr{\"u}ger, N~Kuiger, and Christoph Von Der~Malsburg.
\newblock Face recognition by elastic bunch graph matching.
\newblock {\em IEEE Transactions on Pattern Analysis and Machine Intelligence},
  19(7):775--779, 1997.

\bibitem{yartseva2013performance(30)}
Lyudmila Yartseva and Matthias Grossglauser.
\newblock On the performance of percolation graph matching.
\newblock In {\em Proceedings of the First ACM Conference on Online Social
  Networks}, pages 119--130, 2013.

\bibitem{yu2021power}
Liren Yu, Jiaming Xu, and Xiaojun Lin.
\newblock The power of d-hops in matching power-law graphs.
\newblock In {\em Abstract Proceedings of the 2021 ACM SIGMETRICS/International
  Conference on Measurement and Modeling of Computer Systems}, pages 77--78,
  2021.

\bibitem{zafarani2013connecting}
Reza Zafarani and Huan Liu.
\newblock Connecting users across social media sites: a behavioral-modeling
  approach.
\newblock In {\em Proceedings of the 19th ACM SIGKDD international conference
  on Knowledge discovery and data mining}, pages 41--49, 2013.

\bibitem{zhou2018deeplink}
Fan Zhou, Lei Liu, Kunpeng Zhang, Goce Trajcevski, Jin Wu, and Ting Zhong.
\newblock Deeplink: A deep learning approach for user identity linkage.
\newblock In {\em IEEE INFOCOM 2018-IEEE Conference on Computer
  Communications}, pages 1313--1321. IEEE, 2018.

\end{thebibliography}

\end{document}